\documentclass[twocolumn,showpacs,preprintnumbers,amsmath,amssymb,aps]{revtex4-2}
\usepackage{graphicx}
\usepackage[english]{babel}
\usepackage{dcolumn}
\usepackage{bm}
\usepackage{xfrac}
\usepackage{xcolor}
\setcitestyle{numbers, square}

\begin{document}

\title{Ionization clamping in ultrafast optical breakdown of transparent solids}

\author{Anton Rudenko$^{1,2}$,
Jerome V. Moloney$^{1,2}$, 
and Pavel Polynkin$^{1}$}
\affiliation{$^{1}$College of Optical Sciences, University of Arizona, Tucson, Arizona 85721, USA \\
$^{2}$Arizona Center for Mathematical Sciences, University of Arizona, Tucson, Arizona 85721, USA }
\date{\today}
 
\email{antmipt@gmail.com}
\email{ppolynkin@optics.arizona.edu}
           
\begin{abstract}
We formulate a multi-physics model describing the nonlinear propagation of a femtosecond, near-infrared, tightly focused laser pulse in a wide-band dielectric. 
The application of our model to the case of bulk sapphire shows that even under extreme excitation conditions, ionization is rigidly clamped 
at about one tenth of the electron density in the upper valence band. The earlier estimate of $\sim$10\,TPa pressure 
that could be attainable through the internal excitation of transparent dielectrics by tightly focused ultrafast laser beams is shown to be off by two orders of magnitude.
\end{abstract}
                    
\maketitle


\section{Introduction}

Internal modifications of transparent dielectrics by intense, ultrashort laser pulses find applications 
in diverse areas such as writing optical waveguides, microfluidics, and 3D optical data storage
\cite{gattass2008}. 

When a single femtosecond laser pulse with the energy exceeding a certain threshold 
(typically on the order of 100\,nJ)
is tightly focused below the surface of a transparent solid, rapid deposition of the electromagnetic energy in the material
drives a microexplosion, resulting in the production of a microscopic void \cite{Glezer1996}. 
These voids have been demonstrated in various kinds of transparent solids -- crystals, glasses, and plastics
\cite{mazur1997}, as well as on semiconductor-dielectric interfaces \cite{rapp2015}. 
The generation of extended cylindrical voids inside transparent solids using the excitation with ultrafast Bessel beams
has also been reported \cite{bhuyan2014, Rapp2016, Bhuyan2017}. 

Microexplosion voids are surrounded by shells of densified matter that can contain exotic 
elemental phases \cite{vailionis2011,gamaly2013}.
The mechanism of formation of these super-dense materials, put forth in \cite{gamaly2006,gamaly2013}, 
involves a complete ionization of electrons in the upper valence band of the dielectric,
followed by the local thermalization of the conduction-band electrons with the ions, and
the subsequent kinetic separation of different elements in the expanding solid-density plasma, according to their ionic masses.
It has been argued that the transient pressure attainable inside the microexplosion exceeds 10\,TPa \cite{juodkazis2006}, which is of the same order 
as the pressure inside cores of large planets 
and is comparable to the peak transient pressures that can be realized through laser-driven shocks 
at the national-scale laser facilities \cite{smith2014}. 
        
Computing the distribution of energy deposited by the laser field into the material, 
under the conditions of the laser-driven microexplosion experiments,
is challenging. The early treatments \cite{gamaly2006,gamaly2013} were based on the assumption that 
tight linear focusing of the laser beam dominates over the rest of the beam-shaping mechanisms,
essentially assuming linear propagation of the laser pulse through the interaction volume.
As we will show below, such a treatment leads to the gross overestimation of the major physical quantities
-- the peak laser fluence, plasma density and temperature, and, as a result, the maximum attainable pressure.

The proper account for the complex spatiotemporal dynamics of the laser pulse 
on its highly nonlinear propagation through the interaction volume
necessitates the use of the vectorial Maxwell propagator,
initiated by the non-paraxial input field 
and coupled with a comprehensive material response model accounting for ionization and plasma dynamics. 
Envelope, paraxial, or unidirectional treatments are not adequate for this problem.
Previously reported computational investigations using vectorial Maxwell propagator for the optical field 
limited the tightness of focusing
by considering values of the numerical aperture (NA) in the range 0.3--0.8 and/or limited the input laser fluence,
so that the density of the deposited electromagnetic energy was insufficient
for the generation of an internal void   
\cite{popov2011,bulgakova2015,naseri2020}.
Another simplification used previously was based on the reduction of the dimensionality of the problem to two \cite{Hallo2007,thiele2016}.
In the case of the excitation with a Bessel beam, the approximate translational invariance of the problem allows one
to limit the computational domain along the propagation direction by one optical phase period \cite{morel2022}.
None of those treatments 
is suitable for the quantitative description of the laser-beam propagation under the conditions 
of the representative experiments \cite{juodkazis2006}.        

The use of the 3D Maxwell propagator, which is essential to the problem we consider, is very computationally expensive.  
Formulating the accompanying material response model 
that is both adequate and computationally feasible faces principal challenges.
The treatments based on the kinetic equation for the conduction-band electrons \cite{kaiser2000} 
or those involving multiple energy levels in the conduction band \cite{rethfeld2004} would render the propagation model computationally
prohibitive at the spatial scales of interest.
Only a concise model involving rate equations for the averaged quantities describing the conduction-band electrons
and, possibly, holes, can be practical. 
Since high degree of ionization on 
the order of ten percent or higher is expected,
using constant values for the material parameters that depend on the density of the valence-band electrons, 
such as linear and nonlinear susceptibilities of the crystal host, as well as the energy bandgap,
may be no longer justified. Ideally, those quantities should be parameterized to account for the significant depletion of the valence band,
but such a parametrization would necessarily involve questionable assumptions, compromising the quantitative validity 
of the results. We address this principal hurdle by varying the uncertain parameters, that enter our material model,
in a large range and showing that our conclusions are largely insensitive
to those vast parameter variations.

Our results show that the degree of ionization in sapphire, the representative material used in the confined microexplosion experiments, 
is limited to about 10\% of electrons in the upper valence band
and achieving complete ionization, 
i.e., promoting all valence-band electrons to the conduction band using a multi-cycle, 
near-infrared laser pulse, is not possible. 
Based on our results, we conclude that the earlier estimate
of the $\sim$10\,TPa transient pressure that could be attainable inside a confined microexplosion \cite{juodkazis2006}
is off by about two orders of magnitude.
We suggest potential routes towards overcoming this limitation.

\section{Multi-physics model for the laser-material interaction}
\label{model}

Our consideration focuses on the microexplosions in sapphire, although the treatment can be straightforwardly applied 
to any wide-band dielectric by the appropriate modification of the material parameters.
We assume that the electrons that are available for the transfer to the conduction band via strong-field and collisional ionization 
are all electrons that initially occupy the upper valence band of sapphire, i.e., four {\it 2p} electrons per oxygen atom and three {\it 3p} electrons
per aluminum atom, with the total of 18 electrons per an Al$_{2}$O$_{3}$ ``molecule",
corresponding to the upper valence-band electron density $N_{0} = 4.2 \times 10^{29}$\,m$^{-3}$.
Note that vastly different values for the density of valence-band electrons available for optical ionization in sapphire
had been used in the literature, ranging from one electron per Al$_{2}$O$_{3}$ ``molecule" 
to the complete electron population of the upper valence band, which is the value we use here.

The velocity of an individual conduction-band electron is a vectorial sum of its collective (drift) and random components. 
The drift velocity $\vec{v}_{D}$ determines the macroscopic electron current that drives the optical field, while
the random component is related to the electron temperature. 
The latter is a well-defined quantity due to the very fast (significantly sub-cycle) thermalization of the electron system
by the electron-electron collisions.
The velocity distribution of the conduction-band electrons is approximately Maxwellian, with its center of mass shifted 
by the time-dependent drift velocity of the conduction-band electrons that collectively quiver in the optical field.

To account for the depletion of the upper valence band,
the strong-field and avalanche ionization rates are scaled by the depletion factor 
$\left( 1 - \sfrac{N_{e}}{N_{0}} \right)$, where $N_{e}$ 
is the time- and position-dependent density of the conduction-band electrons. 
We further assume that the linear and third-order susceptibilities of the sapphire host are entirely due to the electrons in the upper valence band 
and that both susceptibilities are linearly proportional to the depletion factor. 
Both linear and nonlinear dielectric responses of the host are assumed to be instantaneous and dispersionless.
Weak birefringence of sapphire is negligible on the spatial scale of the problem we consider. 
Recombination is neglected as it occurs on the multi-picosecond or longer time scale \cite{puerto2010, Bhuyan2017}.
The formation of the self-trapped excitons \cite{TsaiPRL1991} is negligible in sapphire, but it can be straightforwardly 
accounted for in the materials where that effect is appreciable, e.g., in fused silica \cite{Guizard1996}.

The system of equations comprising our model reads:
\begin{eqnarray}
\lefteqn{}
&& \frac{\partial{(\epsilon\vec{E})}}{\partial{t}}
		= \frac{\vec{\nabla}\times\vec{H}}{\epsilon_0}-\frac{1}{\epsilon_0} (\vec{J}_{e} + \vec{J}_{PI}+\vec{J}_{Kerr} ) \, , \nonumber \\
&& \frac{\partial{\vec{H}}}{\partial{t}} = -\frac{\vec{\nabla}\times\vec{E}}{\mu_0} \nonumber \, , \\
&& \frac{\partial{\vec{J}_{e}}}{\partial{t}} 
	= - \nu_e{\vec{J}_{e}} + \frac{e^2N_e}{m_e^*}\left(\vec{E}+ \mu_{0}  \, \vec{v}_{D}\times\vec{H}\right) \nonumber \\
&& \, \qquad \,\,\,\,\,\, - \vec{\nabla}\cdot\left(\vec{J}_{e}\otimes\vec{v}_{D}\right) 
	- \frac{e}{m_{e}^*} \vec{\nabla} P_{e} \nonumber \, , \\
&& \frac{\partial{N_e}}{\partial t}  
	= \left( 1 - \sfrac{N_e}{N_0} \right) \left(w_{PI} + w_{AI}\right) -  \vec{\nabla}\cdot\left(N_e\vec{v}_{D}\right) \nonumber \\
&& \, \qquad \,\,\,\,\,	+ D_a \Delta{N_e} \nonumber \, , \\ 
&& \frac{\partial{\mathcal{E}_e}}{\partial t}  
	= ( \vec{J}_e+\vec{J}_{PI} ) \cdot\vec{E} - \vec{\nabla}\cdot\left[\left(\mathcal{E}_e + P_e\right)\vec{v}_{D}\right] \nonumber \\
&& \, \qquad \,\,\,\,\, + \vec{\nabla}\cdot\left(\kappa_e \vec{\nabla}{T_e}\right)  ,
\label{Plasma}
\end{eqnarray}
where  $\vec{E}$ and $\vec{H}$ are electric and magnetic fields, and the SI units and the standard notation for the basic constants are used throughout.
The linear dielectric response of the valence-band electrons is described by the dynamic susceptibility $\epsilon$, while the contributions from the plasma, ionization,
and Kerr effect are accounted for through the corresponding currents in the equation for the $\it E$-field. Explicitly:
\begin{eqnarray}
&& \epsilon = 1 + (n_{0}^{2}-1)(1-N_e/N_0) \, , \label{epsilon} \\
&& \vec{J}_{e} = e{N_e}\vec{v}_D \, , \label{J_e} \\
&& \vec{J}_{PI} = w_{PI}I_p\frac{\vec{E}}{\vert \vec{E} \vert ^{2}}(1-\sfrac{N_e}{N_0}) \, , 
\label{J_Kerr} \\
&& \vec{J}_{Kerr} = \frac{4n_{0}^2n_{2,0}\epsilon_0^{3/2}}{3\sqrt{\mu_0}}\frac{\partial}{\partial{t}}\left[(1-\sfrac{N_e}{N_0})\vec{E}(\vec{E}\cdot\vec{E})\right]  ,
\end{eqnarray} 
where, for the unperturbed sapphire, the linear refractive index $n_{0} = 1.76$ 
and the nonlinear Kerr index $n_{2,0} = 3\cdot{10}^{-20}$ m$^2$/W \cite{Major2004}. 
$T_{e}$ is the electron temperature (the measure of energy 
associated with the random motion of the conduction-band electrons),
and $I_p = 9.9$ eV is the ionization potential, a constant value in our model.
$m_e^*$ is the effective mass of the conduction-band electrons, with the value of about $0.4\,{m_e}$, where $m_e$ is the free-electron mass \cite{medvedeva2007}. 
$\nu_e(N_e, T_e)$ is the effective collision rate of the conduction-band electrons with massive particles, that facilitate electron heating by the laser field; 
different types of collisions dominate over different ranges of electron density and temperature, as detailed in Appendix \ref{appendix3}. 
The contribution to the macroscopic current by holes is neglected, since they are about ten times heavier 
and, correspondingly, less mobile than the conduction-band electrons \cite{medvedeva2007}. 
The last term in the rate equation for $\vec{J}_e$ is due to the divergence of the electron pressure $P_e = N_e{k_B}T_e$. 
Assuming the adiabatic equation of state with the adiabaticity constant $\gamma = 5/3$ for the electron gas, it is evaluated as 
$\vec{\nabla} P_{e} = \left( 5 / 3 \right)  k_{B} T_e \vec{\nabla} N_{e}$ \cite{lieberman2005}.

Electrons in the conduction band are generated via Keldysh photo-ionization and avalanche ionization mechanisms,
at the rates $w_{PI}$ and $w_{AI}$, respectively; these rates are detailed in Appendices \ref{appendix1} and \ref{appendix2}.
The photo-ionization rate is evaluated at the fixed, center frequency of the incident laser pulse. 
The avalanche ionization rate is derived by averaging the rate for a single impactor electron over the velocity
distribution of the conduction-band electrons \cite{Penano2005}.
That velocity distribution is assumed to be Maxwellian with its center of mass shifted by the drift velocity of the collective motion 
of the conduction-band electrons, which is directly related to the macroscopic current 
(see Appendix \ref{appendix2} for details).
In this formulation, the impact ionization continues after the passage of the laser pulse, through the utilization of the hot electrons 
left in its wake. By energy conservation, that after-burn ionization process does not affect the distribution of the total deposited energy. 

The hydrodynamic terms in the last three equations in (\ref{Plasma}) account, in the standard way, for the convection,
diffusion, which we assume to be ambipolar, and thermal transport in the conduction-band electron system.
The ambipolar diffusion coefficient, under the conditions of the electron temperature and mobility being much higher 
than the corresponding values for the ions, is given by $D_a = \mu_{i}k_{B}T_{e}/e$, where $\mu_i = {e}D_i/k_{B}T_i$ \cite{lieberman2005}, 
and $D_i = \kappa _{0} / \rho \, C_0 = 1.13\cdot{10}^{-5}$ m$^2$/s are the ionic mobility and the thermal diffusivity for sapphire 
at $T_i = 300$ K (the temperature of the cold lattice), respectively. 
In  the above expressions, $C_0 = 780$\,J/(kg$\cdot$K) is the heat capacity \cite{Ci},
$\kappa _{0} = 35$\,W/(m$\cdot$K) is the thermal conductivity \cite{kappa}, and $\rho = 3,980$\,kg/m$^3$ is the density of sapphire, also all at room temperature. 
The electron thermal conductivity is defined as $\kappa_{e} = 2k_B^2{\mu_e}{N_e}{T}_e/e$, 
where the electron mobility $\mu_e = 3\times{10}^{-5}$\,m$^2$/(V$\cdot$s) \cite{Bulgakova2010}.

We point out that a significant uncertainty of the
transport constants, used here, exists, and that those values are dynamic, i.e., depend on the level of excitation.
However, as our simulations show, among all hydrodynamic terms, only the electron diffusion has a non-negligible effect on the laser-energy deposition, 
and its contribution is limited to $\sim$20\% of the total. Thus, the uncertainty 
in the transport constants does not significantly affect our conclusions.

The instantaneous volumetric density of energy stored in the electron subsystem
is composed of the thermal, drift, and ionization components:
\begin{eqnarray}
\mathcal{E}_e = \left(\frac{3}{2}k_B{T_e}+\frac{m_e^*{v_D^2}}{2}+I_p\right){N_e}
\label{energy}
\end{eqnarray}
In our simulations, the current values of $\mathcal{E}_e$, $\vec{J}_{e}$, and $N_{e}$, 
computed by integrating equations (\ref{Plasma}),
are used to determine the value of $T_{e}$ from (\ref{energy}). 
$T_{e}$ is then used to update the ionization and transport constants. 
Our goal is to compute the distribution of the deposited energy (\ref{energy}) immediately after the passage of the laser pulse through the interaction volume,
at which point the drift contribution to the energy vanishes, as the conduction-band electrons are no longer driven by the optical field.

In order to implement the significantly non-paraxial focusing of the incident laser beam 
in the Finite-Difference Time-Domain (FDTD) solver for the system of equations (\ref{Plasma}), the boundary conditions representing
the electromagnetic field source at the entrance boundary
of the computational domain need to be set appropriately and self-consistently with the Maxwell equations.
For large values of the numerical aperture (NA\,$\gtrsim 0.3$), 
the paraxial approximation, in which only two components of the incident field (e.g., $E_{x}$ and $H_{y}$)
on the entrance boundary are specified, fails, as discussed in Refs. \cite{bulgakova2014, thiele2016}. 
All three components of both $\vec{E}$ and $\vec{H}$ field vectors need to be specified such that the fields
satisfy the Gauss law (i.e., have zero divergence). 
The proper way to set the boundary conditions for the field vectors consistently with the experimental 
realization is through the exact solution of the linear propagation of the incident field that 
is focused by a high-NA focusing optic, e.g., a parabolic mirror.
Such a solution has been formulated by Stratton and Chu in the form of diffraction integrals \cite{stratton1939}.    
For the case when the focal length of the focusing optic is much larger
than the wavelength, those integrals can be written in a concise form referred to as the Ignatovsky representation.
In our formulation of the boundary conditions, we follow Peatross et al. \cite{peatross2017}.
The explicit formulas are given in Appendix~\ref{boundary}, for completeness.

Our simulation computes the distributions of electric and magnetic fields inside and around the interaction volume. 
Fluence cannot be correctly defined for a non-paraxial field.
Nevertheless, to express our results in the units of that familiar quantity, 
we introduce the fluence through its generic definition,
strictly valid only for plane waves: $F \equiv \int c n_{0} \epsilon _{0} \vert \vec{E} \vert ^{2}  \, dt$, 
where $\vec{E}$ is the instantaneous $E$-field (not the field envelope, thus there is no factor of 1/2 in the definition) 
and $n_{0} = 1.76$ is the refractive index of unperturbed sapphire. 
The integration is over the time interval covering the optical pulse.    

The numerical scheme for solving the system of the 3D Maxwell equations with the linear and nonlinear source terms, comprising our model,
is based on the Finite-Difference Time-Domain (FDTD) method commonly used for solving linear Maxwell equations \cite{Yee1966,Inan2011} 
and a fixed-point iteration algorithm for evaluating the source terms \cite{Rudenko2018_GPU}. The details of the numerical scheme and the computational aspects
are discussed in Appendix~\ref{numerical}.

The conceptual weakness of our model, just like of many other models based on the rate equations for the material response to a strong optical field, 
is in the generic treatments of the strong-field and avalanche ionization mechanisms, 
as well as of the electron heating. The rates of those complex processes are expressed in terms of basic constants, with the valence-band electron density, 
the upper edge of the energy bandgap, and the effective electron mass being the only material parameters used.
To address this principal hurdle, we have independently varied the uncertain quantities that enter our model 
in large ranges of up to one order of magnitude up and down from their baseline values collected from literature. 
The details of this analysis are given in Appendix \ref{parameters}.
Those results show that vast variations of the ionization and collision rates result in only moderate changes of the maximum value 
of the deposited energy density and its spatial distribution. Plasma shielding,
which is the mechanism limiting ionization, is robust and essentially independent of the specifics of the wide-band dielectric that is excited by the tightly focused, 
near-infrared laser pulse.

\section{Estimation of temperature and pressure of sapphire in the WDM state}

Immediately after the electromagnetic energy is deposited into the conduction-band electron system by the laser pulse, 
the sample is left in the non-equilibrium state comprised of hot electrons and cold crystal lattice, still maintaining its spatial periodicity.
The hot electron gas has pressure $p_{e} = N_{e} k_{B} T_{e}$. It can be high, but it is not the ``operative" pressure
driving extreme phase transformations, as the system exists in this non-equilibrium state for a short period of time, significantly before the void starts to open. 
Within ten to one hundred picoseconds after the passage of the laser pulse, the electron energy is transferred to the sapphire host,
and the local thermal equilibrium (LTE) state is reached, where the electrons and the host are at the same, position-dependent, temperature,
and the host is no longer spatially periodic. The system is in the warm dense-matter (WDM) state, characterized by the LTE temperature and extreme pressure,
which drives the transformations to the super-dense material phases observed experimentally \cite{vailionis2011}.
The electrons-to-lattice energy transfer involves various interaction channels.
The details of this process can be complex but they are not important for the purposes of estimating the peak temperature and pressure of
sapphire in the WDM state, as long as the LTE state is reached before the microexplosion void starts to open, which is the assumption we follow.
To estimate the peak temperature of the WDM sapphire, we use the SESAME isotherms \cite{sesame,sesame1} that quantify the internal energy of sapphire, 
estimated as $\mathcal{E} _{e} (t=600$\,fs$)$ and converted to MJ/kg, at its solid, unperturbed density of $\rho = 3,980$\,kg/m$^{3}$.
Then, using thus found value of the peak LTE temperature, we use the other form of the SESAME equation of state
for sapphire, also graphed in \cite{sesame,sesame1}, to find the pressure of sapphire at that temperature, and, again, at its solid, unperturbed density.    
This is the ``operative" pressure behind the extreme phase transformations. It can be used as an input to the hydrodynamic codes that simulate
the dynamics of the formation of the microvoid on the hundreds of picoseconds to the nanoseconds time scales \cite{Mezel2008}.

For example, the peak deposited volumetric energy of 300\,nJ/$\mu$m$^{3}$ deposited in the solid-density sapphire,
corresponds to the energy density, per unit mass, of about 75\,MJ/kg. From the SESAME isotherms obtained from the equation of state \#7411
\cite{sesame,sesame1}, we find that the temperature of the WDM sapphire, corresponding to that value of deposited energy, is about 3.5$\times$10$^4$\,K
(isotherm M). The pressure, corresponding to that temperature, also at the solid density, is estimated at 250\,GPa.  

\begin{figure}[b!]
\includegraphics[width=90mm]{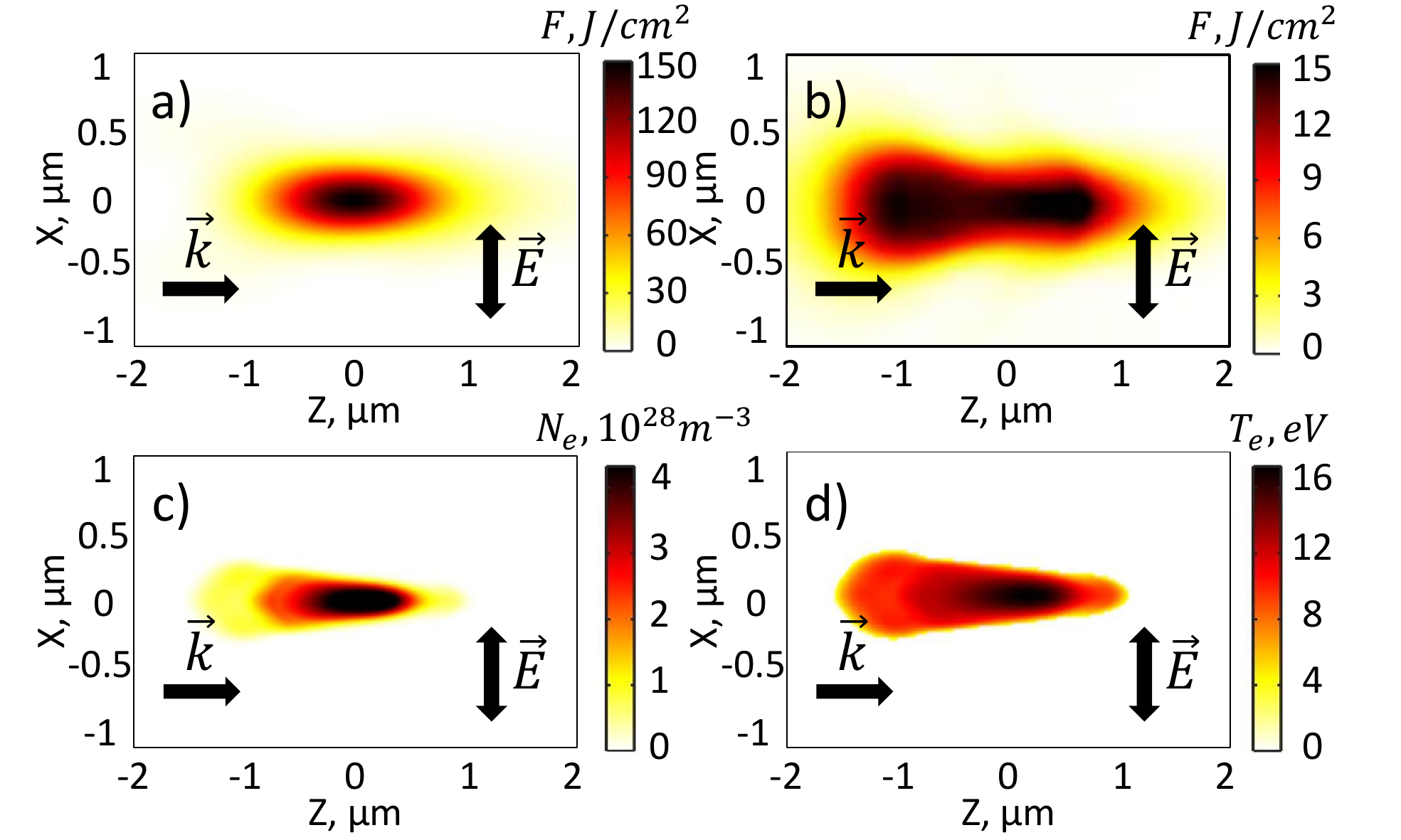}
\caption{\label{fig_1m} XZ-slices of the spatial distributions of (a) laser fluence, defined in the text, for the laser pulse propagating linearly, (b) the same for the nonlinear propagation, 
(c) density and (d) temperature of the conduction-band electrons at 300\,fs after the the passage of the peak of the laser pulse through the position of the linear focus.
The incident laser pulse with the center wavelength of 800\,nm has the FWHM duration of 150\,fs and energy of 100\,nJ;
the effective NA of the focusing optic is 1.35. The incident laser pulse is linearly polarized along the X-axis and propagates along the Z-axis.}
\end{figure}

\section{Results and discussion}

Fig.~\ref{fig_1m} summarizes the simulation results for the representative experimental realization of the laser-driven microexplosion in sapphire \cite{juodkazis2006}.
The excitation conditions are specified in the figure caption. The distribution of the optical fluence, as defined above 
through the computed electric field,
for the case of linear propagation (i.e., with all terms accounting for ionization and the nonlinear response dropped from the equations) is shown in the panel (a),
while the corresponding distribution for the nonlinear case is shown in the panel (b).
According to these simulations, ionization and the effects associated with it reduce the peak fluence
by about one order of magnitude relative to the case of linear propagation. 
The distributions of the density and temperature of the conduction-band electrons left in the wake of the laser pulse, 
shown in the panels (c) and (d), respectively, reveal the formation of an asymmetric plasma structure
with the maximum values reached approximately in the center of the focal volume. 
The maximum value of the free-carrier density is about $4 \times 10^{28}$ m$^{-3}$, which is by one order of magnitude 
lower than the total electron density in the upper valence band of sapphire. 
The peak thermal energy per one conduction-band electron at the focus is about 15\,eV, which is 1.5 times the value of the band gap.

\begin{figure}[t!]
\includegraphics[width=90mm]{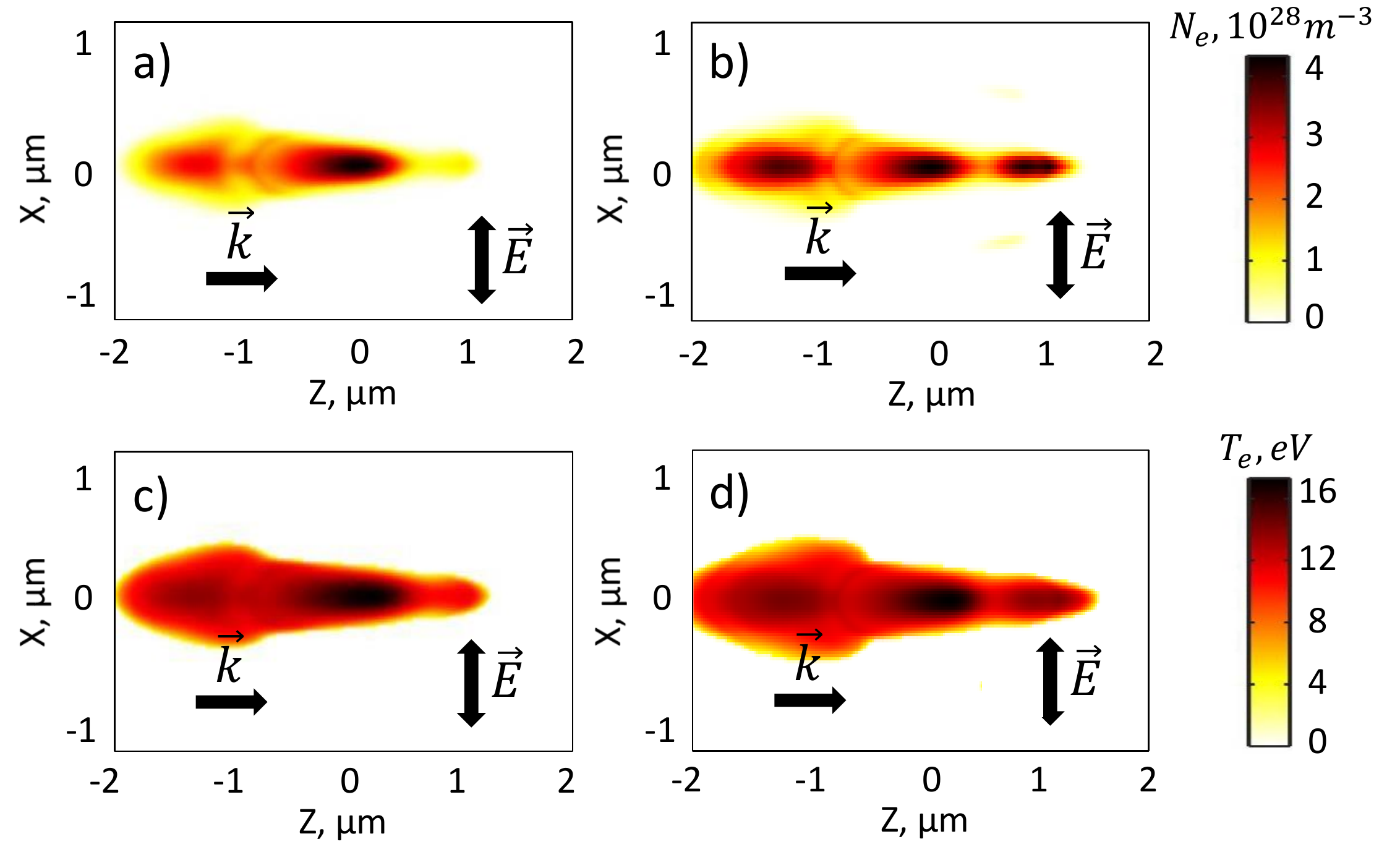}
\caption{\label{fig_2m} Spatial distributions of (a,b) the conduction-band electron density and (c,d) temperature
for the cases of input pulse energies of (a,c) 400\,nJ and (b,d) 1\,$\mu$J. 
The other conditions are the same as those in Fig.~1. 
Additional plasma balls are formed before and after the focus, while the peak electron density and temperature remain
rigidly clamped.}
\end{figure}

Increasing the incident pulse energy does not lead to the appreciable enhancement of either the peak density or the peak temperature 
of the plasma, as demonstrated by the simulation results shown in Fig. \ref{fig_2m}. Here, we used the input pulse energies four and ten times
the value used in the case shown in Fig. \ref{fig_1m}, while keeping the focusing conditions unchanged. The difference between these 
cases and the one with the lower input pulse energy is that the impenetrable plasma skin layer in the middle 
of the focal volume is formed earlier in the pulse. The energy left in the trailing edge forms additional plasma balls before and after the focus.
We point out that the rotational asymmetry of the distributions, shown in Figs. \ref{fig_1m} and \ref{fig_2m}, around the Z-axis is non-vanishing but insignificant. 

\begin{figure}[t!]
\includegraphics[width=90mm]{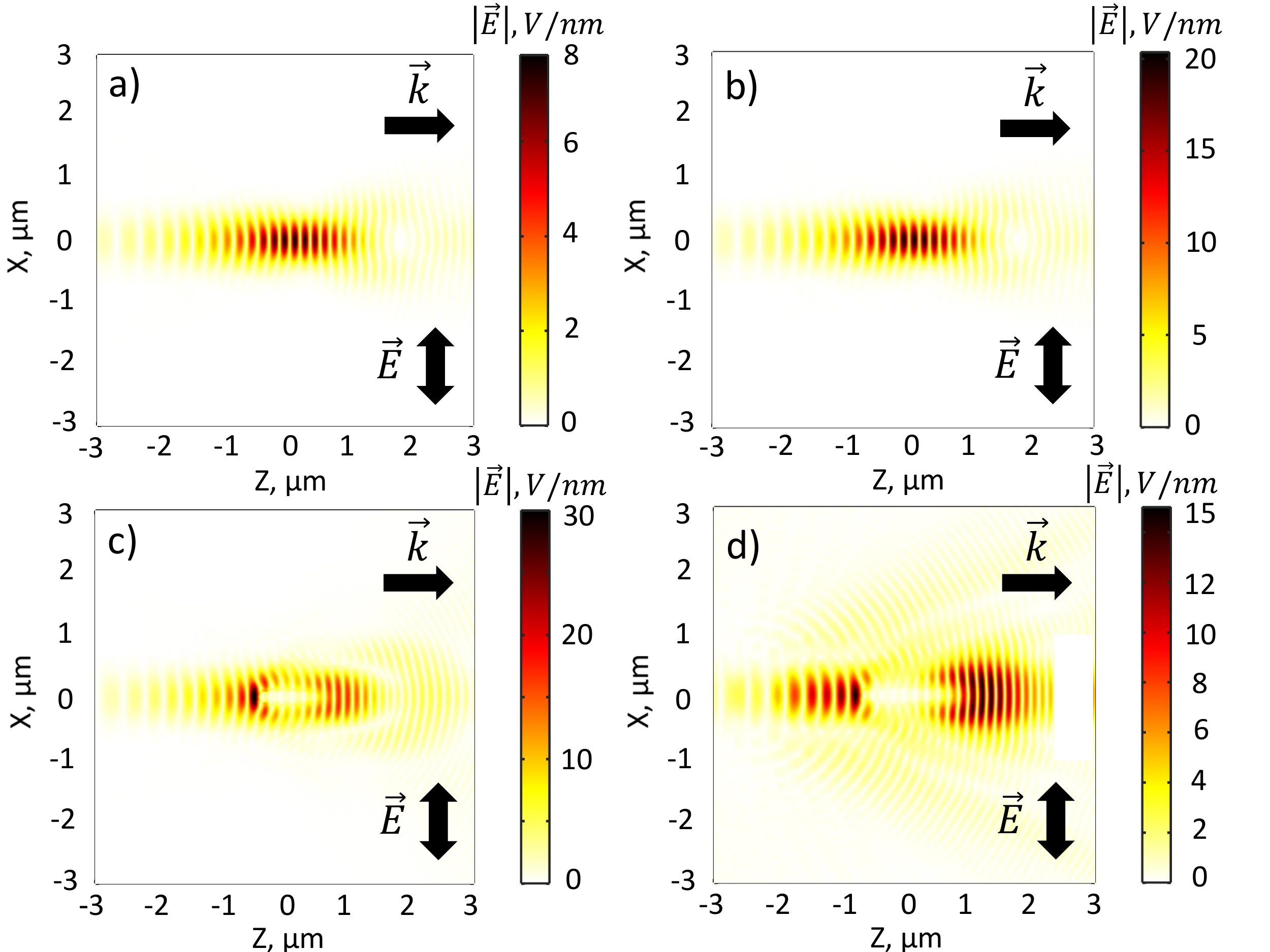}
\caption{\label{fig_7}
Instantaneous electric field distributions: (a) at $120$\,fs before the pulse peak reaching the focus, (b) at $60$\,fs before the pulse peak reaching the focus, 
(c) with the pulse peak at the focus, and (d) at $60$\,fs after the pulse peak passing through the focus.
The energy of the incident laser pulse is 100\,nJ. 
The focusing conditions and other parameters are the same as those used in Figure~\ref{fig_1m}.}
\end{figure}

\begin{figure}[b!]
\includegraphics[width=90mm]{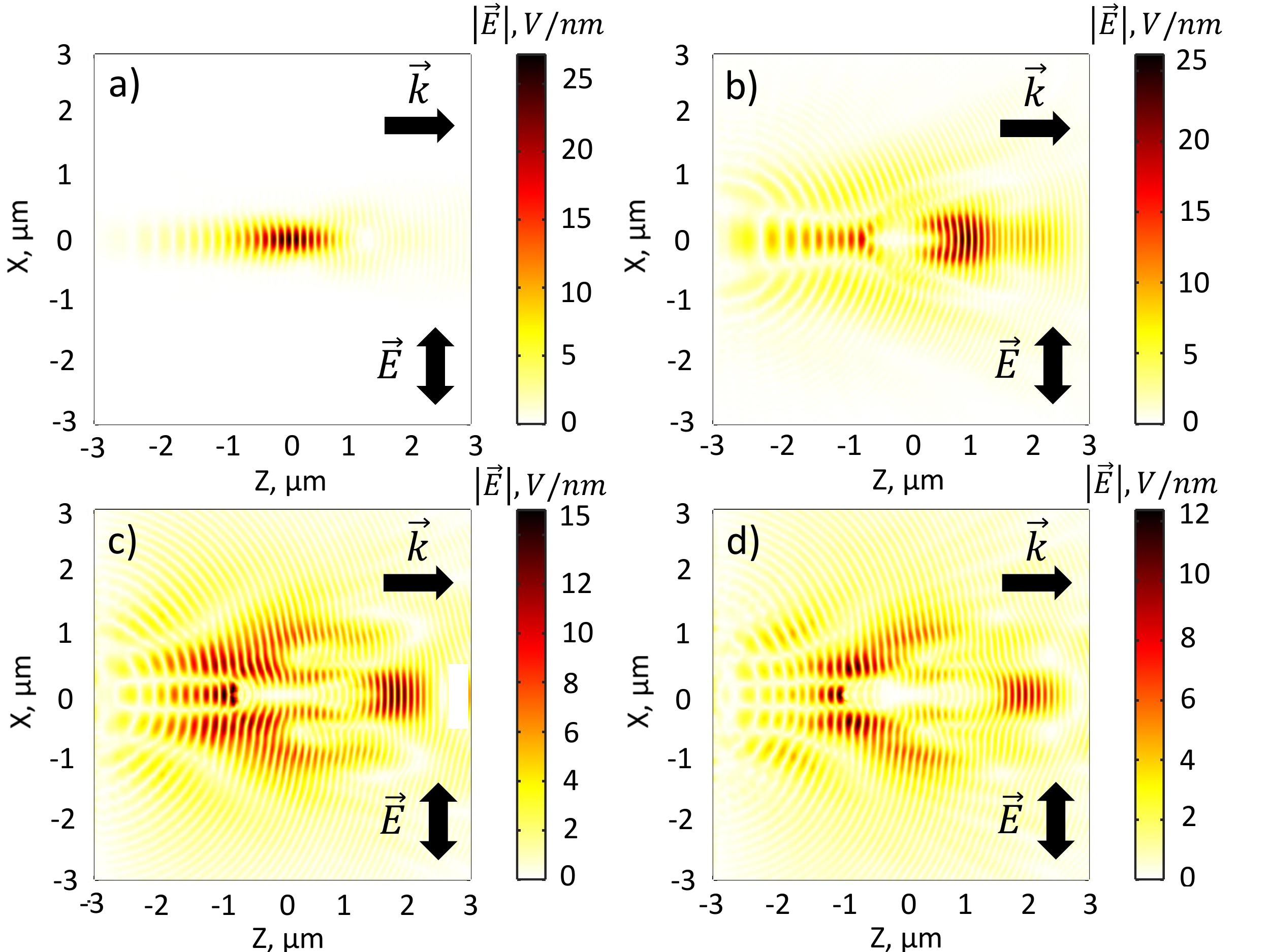}
\caption{\label{fig_8}
Instantaneous electric field distributions: (a) at $120$\,fs before the pulse peak reaching the focus, (b) at $60$\,fs before the pulse peak reaching the focus, 
(c) with the pulse peak at the focus, and (d) at $60$\,fs after the pulse peak passing through the focus.
The energy of the incident laser pulse is 1\,$\mu$J. 
The focusing conditions and other parameters are the same as those used in Figure~\ref{fig_1m}.}
\end{figure}

\begin{table*}[tb!]
\centering
\begin{tabular}{|c||c|c|c|c|c|} 
\hline Material parameters & Fluence, J/{cm}$^2$ & $ N_e$, $10^{28}$ m$^{-3}$ & $T_e$, eV & Energy density, nJ/$\mu$m$^3$ & Peak WDM pressure, GPa \\
\hline
\hline Nominal values ($\times{1}$) & 7.5 & 4.5 & 15 & 235 & 200 \\
\hline ${0.1}\times{w_{PI}}$ & 10 & 6 & 12.5 & 275 & 225 \\
\hline $10\times{w_{PI}}$ & 7.5 & 3 & 22 & 205 & 175 \\
\hline ${0.1}\times{w_{AI}}$ & 12 & 1.5 & 65 & 260 & 225 \\
\hline $10\times{w_{AI}}$ & 6 & 8 & 9 & 300 & 250 \\
\hline $\nu_e = 10^{15}$ s$^{-1}$ & 7.5 & 2.5 & 31 & 225 & 200 \\
\hline $\nu_e = 5\cdot{10}^{15}$ s$^{-1}$ & 6 & 3 & 17 & 170 & 150 \\
\hline $\tau_{rec} = 1$ ps & 7.5 & 3 & 24 & 220 & 200 \\
\hline $\tau_{rec}(N_e)$ & 7.5 & 1.5 & 60 & 240 & 200 \\
\hline \end{tabular}
\caption{\label{table1} 
Summary of the simulation results with varying values of the system parameters.} 
\end{table*}

It is instructive to trace the dynamics of the laser pulse on its propagation through the interaction volume for different values of the input pulse energy. In Figures~\ref{fig_7} and \ref{fig_8}, we show the snapshots of the instantaneous electric field distributions 
for the input pulse energies of 100\,nJ and 1\,$\mu$J, respectively, at different moments of time.
In both cases, the field flows around the over-critical plasma ball formed in the middle of the focal volume, without penetrating it and without enhancing the peak intensity above the value of about $2\times$10$^{18}$\,W/m$^{2}$ (corresponding to the peak electric field of 
$\sim 30$\,V/nm). In the case of the higher input pulse energy, the trailing edge of the pulse flows around the center of the focal volume and creates additional intense field concentrations before and after the focus.

The ionization clamping effect sets a limit on the volumetric density of energy deposited into the medium by the laser field. 
From the simulations, the peak free-electron density  
immediately after the passage of the laser pulse through the interaction zone is 4$\times$10$^{28}$\,m$^{-3}$, 
while the peak value of the electron temperature is about 15\,eV. 
The gap contribution to the deposited energy accounts for additional $\sim$10\,eV per one conduction-band electron.
The total peak density of energy deposited into the medium is 
$\mathcal{E}_e ^{max} = \left[ \left( 3/2 \right) k _{B} T_{e} + I_{p} \right] N_{e} \approx 235$\,nJ/$\mu$m$^{3}$.

Within few picoseconds after the excitation, the conduction-band electrons thermalize with the sapphire host.
The system reaches the warm dense matter (WDM) state
before the void starts to open \cite{gamaly2006}, 
i.e., while the density of the material throughout the interaction volume remains at its value for the unperturbed sapphire. 
Based on the SESAME equations of state for the energy and pressure of sapphire under extreme conditions \cite{sesame,sesame1}, the value of the peak density of absorbed energy corresponds to the maximum temperature and pressure of the WDM sapphire of $\sim$3$\times 10^{4}$ K and $\sim$200\,GPa, respectively.
The peak temperature is by one order of magnitude and the peak pressure - by two orders of magnitude lower
than the corresponding values purported in \cite{juodkazis2006}.
Note that the peak estimated pressure is about one half of the Young's modulus of the unperturbed sapphire (400\,GPa),
yet it is sufficiently high for producing an internal void within the crystal and causing extreme material phase transformations.

\section{Robustness of the results against variations of material parameters}

We have thoroughly checked the robustness of our results against the potential uncertainty 
of the ionization and electron heating rates. These checks, detailed in Appendix \ref{parameters}, are summarized in Table~I. 
All values shown are the maxima over the interaction volume.
A FWHM 150\,fs pulse with the center wavelength of 800\,nm and pulse energy of 100\,nJ,
focused into the bulk of sapphire with the effective NA=1.35, is used in all cases.  The varied parameter, a single one in each simulation, is specified in the leftmost column. 

The top row summarizes the results obtained using the nominal parameter values and neglecting electron recombination.   
Note that the deviations of the peak electron density and temperature from their values, obtained with the nominal parameters, 
are typically in opposition to each other, so that the corresponding variations of the peak deposited energy density and the peak pressure 
of sapphire in the WDM state are fractionally lower than the variations of either density or temperature.

In summary, over the entire range of parameter variations that we considered, the maximum density of deposited energy and the peak pressure 
of sapphire in the WDM state are limited to 300\,nJ/$\mu$m$^{3}$ and 250\,GPa, respectively.

\section{Potential routes towards overcoming the ionization camping limit}

Due to the scaling of the critical plasma density with wavelength, the excitation with UV and deep-UV laser pulses 
(that have to remain within the linear transparency window of the host dielectric) 
may push the clamped values higher, making a complete ionization of all valence-band electrons achievable.
An example of the excitation of sapphire with the femtosecond laser pulse at the 400\,nm carrier wavelength is discussed in Appendix~\ref{wavelength}.
The peak conduction-band electron density and temperature are predictably higher than the corresponding values obtained with the excitation 
at the 800\,nm wavelength. However, the peak pressure of the WDM sapphire is still about two orders of magnitude lower than the 
$\sim$10\,TPa value purported in \cite{juodkazis2006}.

Other potential routes towards overcoming the clamping limits discussed here may be through the application 
of single-cycle laser pulses or through the use of spatial and temporal pulse shaping, 
e.g., via simultaneous spatial and temporal focusing (SSTF) \cite{SSTF2005,OL2023}. 
The practicality and limitations of those approaches remain to be investigated. 

\section{Conclusions}

In conclusion, we have applied a fully vectorial, 3D Maxwell propagator, initiated by the non-paraxial input optical field,
to compute the energy deposition by a femtosecond, near-infrared laser pulse in bulk sapphire under very tight focusing conditions. 
Plasma shielding in the interaction volume rigidly clamps
the density and temperature of the conduction-band electrons and limits
the peak pressure in the WDM state of sapphire to about 200\,GPa,
about two orders of magnitude lower than the estimate reported previously, yet sufficiently high to drive extreme 
phase transformations. Our model produces quantitatively consistent distributions of the deposited energy 
under large variations of ionization and electron-heating rates 
and can guide future experimental investigations of extreme phase transformations on the table top.
We have suggested the potential routes towards overcoming the ionization clamping limits in the confined microexplosion experiments.

\begin{acknowledgments}
This work was supported by the US Air Force Office of Scientific Research under awards \#FA9550-12-1-0482, \#FA9550-19-1-0032, \#FA9550-17-1-0246, and \#FA9550-21-1-0463, 
and by the US Office of Naval Research under award \#N00014-21-1-2469. The authors thank Miroslav Kolesik, Anton Gusakov, and Mikhail Ivanov for helpful discussions.
\end{acknowledgments}

\appendix

\section{Formulations for the electron collision and ionization rates}

\subsection{Electron collisions} 
\label{appendix3}

In the Drude model, collisions of the conduction-band electrons with heavier species mediate the net absorption of the electromagnetic energy by the electrons.
Quantitative description of the relevant collision processes in real solids is a complex and often untraceable problem,
and its treatments differ significantly across published literature. 
In a crystal excited by an ultrashort laser pulse,
the ions remain nearly frozen during the excitation. The conduction-band electrons move freely in the nearly periodic potential 
of the ions and, at least at the relatively low electron density and temperature, do not directly interact with the individual ions 
constituting the crystal lattice. 
What the electrons interact with are: other electrons, lattice defects and impurities, the deviations from the lattice periodicity (i.e., phonons), 
and holes. We assume that the crystal is defect- and impurity-free. The electron-electron collisions are responsible for the rapid
thermalization of the electron system, but they do not contribute to the electron
heating, since they do not change the total momentum of the colliding electrons. Thus, the two collision 
processes that are relevant for the heating of the electron system in our model, at the low-to-moderate level of excitation, 
are the collisions with the phonons and with the holes.

Sapphire has 3 acoustic and 27 optical phonon modes \cite{Al2O3phonons}. A comprehensive description of the electron-phonon collisions 
that accounts for both phonon types has been recently published in \cite{preprint}, where the contributions to the
total electron-phonon collision rate by different phonon modes
have been computed using the electron density of states in the conduction band of sapphire \cite{DOS}.
The result for the total collision rate varies between $\sim1 \times 10^{14}$\,s$^{-1}$ and $\sim5 \times 10^{14}$\,s$^{-1}$
over the range of electron energies from few tens of meV to 5\,eV, where the electron-hole and electron-ion collisions start to dominate.
Accordingly, for the combined electron-phonon collision rate we adopt the average value of $3 \times 10^{14}$\,s$^{-1}$,
independently of the electron energy. Using constant electron-phonon collision rate is common in numerical treatments of ultrafast ablation of dielectrics \cite{waedegaard2013,garcia2017}.   

Holes in sapphire are about ten times heavier than conduction-band electrons \cite{xu1991,medvedeva2007}, thus the electron-hole collisions do contribute to the heating
of the conduction-band electrons by the optical field. 
The rate of the electron-hole collisions, as a function of the electron temperature, is given by 
$\nu_{e-h} = \left( \pi\epsilon_0 / 2e^2 \right) \sqrt{3 / m_e^*}{(k_B{T_e})}^{3/2}$ \cite{ramer2016}. 
This expression is valid if the effective hole mass is much larger than the effective electron mass, which is the case 
for the photo-excited sapphire \cite{medvedeva2007}.
The above expression for the electron-hole collision rate does not depend on the nature of the charged particles and does not account for the effects of the Coulomb long-distance interaction. The latter becomes important for dense plasmas $N_e \gtrsim 10^{27}$ m$^{-3}$,
but there the electron-ion collisions start to dominate, as we discuss next.

It has been argued that for hot electron plasma with the temperature exceeding the Fermi temperature
$T_F = \left( 1/k_{B} \right) \hbar^2{(3{\pi}^2N_e)^{2/3}} / 2m_e^*$, where $N_{e}$ is the electron density, 
the electron-lattice interaction reduces to the Coulomb scattering of the conduction-band electrons on the individual ions constituting the lattice \cite{chimier2011}. The validity of such a treatment is questionable, but we will follow it, keeping in mind that the resulting total collision rate will be varied and our simulation results will be shown to be weakly dependent on those variations.
 
The explicit dependence of the electron-ion collision rate that we use is as follows \cite{eidmann2000}: \\
$\nu_{ei}(N_e, T_e) = (4/3) \sqrt{2\pi}
\left[ e^4{N_e}\ln{(\Lambda)} / \left( \epsilon_0^2\sqrt{m_e^*}({k_B}T_e)^{3/2} \right) \right]$, 
where $\ln{(\Lambda)} = (1/2) \ln\left[1+\left(b_{\max} / b_{\min} \right)^2\right]$ is the Coulomb logarithm, 
$b_{\max} = \sqrt{k_B{T_e}/m_e^*} /\max \left[ \omega _{0}, \omega_{pl} \right]$ and 
$b_{\min} = \max \left[ \left( e^2 / k_B{T_e} \right) , \left( \hbar / \sqrt{m_e^*{k_B}T_e} \right) \right]$ are the maximum and minimum collision parameters, and $\omega_{pl} = \sqrt{e^2 N_e / \left(\epsilon_0 m_e^* \right) }$ is the electron plasma frequency. \\

The upper limit for the electron-lattice collision rate for dense and hot conduction-band electron gas
is set by the shortest possible electron mean-free path within the lattice. It is on the order of the average distance 
between the neighboring ions, which is related to the average atomic number density of sapphire.
This limit is referred to as the natural limit.
In the average sense, neglecting the complex structure of the elementary crystal cell, this limit is approximately given by  
$\nu_{max}(|v_e|) = |v_e|\sqrt[3]{4\pi N_{i,0}/3}$ \cite{eidmann2000, chimier2011, gamaly2014},
where $|v_e|= \sqrt{v_D^2 + 3{k_B}{T_e}/{m_e^*}}$ is the average velocity of the electrons, which includes the drift and thermal components, 
and $N_{i,0} = 1.2\times{10}^{29}$\,m$^{-3}$ is the total ionic number density (aluminum and oxygen ions) of sapphire.
Note that $N_{i,0}$ is different from $N_0$, the density of the valence-band electrons.

In the implementation of the above model for collisions, the rates of the three collision types that are accounted for (electron-phonon, electron-hole, and electron-ion) are computed according to the formulas detailed above for the entire range of the electron density and temperature. The effective collision rate is then calculated as a sum of those three collisions rates, which is the approximation commonly referred to as the Matthiessen's rule
\cite{Matthiessen}. The result of this computation ends up to be $\sim 3 \times {10}^{14}$ s$^{-1}$ at the low level of excitation,
where the electron-phonon scattering dominates, increases with the electron temperature $\propto T_e^{3/2}$ up to $\sim {10}^{15}$ s$^{-1}$ due to the contribution of the electron-hole collisions in the dilute electron plasma, where that process dominates,
and then increases with increasing the electron density $\propto N_e/{T_e}^{3/2}$ due to the electron-ion collisions in
the dense, near-critical or overcritical plasma. 
For a high level of excitation, the electron collision rate is capped at the above-mentioned natural limit, which depends on the local average electron velocity and is $\sim 2.4 \times 10^{16}$~s$^{-1}$ for 10\,eV electrons.

\subsection{Keldysh photo-ionization rate}
\label{appendix1}

Keldysh photo-ionization rate is computed using the following formula \cite{gruzdev2005}:
\, \\
\begin{eqnarray} \lefteqn{} w_{PI}\left(|\vec{E}|, \omega _{0}, I_p \right) = \frac{4\omega _{0}}{9\pi}
\left(\frac{\omega _{0} m_e^*}{\hbar\gamma_1}\right)^{3/2} Q(\gamma,x) \nonumber \\
\times\exp\left[-\pi \cdot \mbox{Int}(x+1) \frac{K(\gamma_1) - M(\gamma_1)}{M(\gamma_2)}\right],
\label{wPI}
\end{eqnarray}
where $\omega _{0}$ is the carrier frequency of the optical field, 
$K$ and $M$ are the complete elliptic integrals of the first and second kind, respectively, 
$\gamma_1 = \gamma/\sqrt{1+\gamma^2}$, $\gamma_2 = 1/\sqrt{1+\gamma^2}$,
$\gamma = \omega _{0}\sqrt{m_e^{*}I_p}/ ( e|\vec{E}| )$ is the Keldysh parameter, 
$I_p$ is the ionization potential (the lower edge of the band gap, a constant value in our model),
$|\vec{E}|$ is the absolute value of the instantaneous electric field,
and $x = 2{I_p}M(\gamma_2) / \left(\pi\hbar\omega_{0} \gamma_1 \right)$. 
The $Q$ integral is defined as follows: 
\begin{align*}
\lefteqn{}
Q(\gamma,x) = \sqrt{\pi/\left( 2 K(\gamma_2) \right)} 
\sum_{n=0}^\infty\left[\exp \left(-\pi{n}\frac{K(\gamma_1)-M(\gamma_1)}{M(\gamma_2)} \right) \right. \nonumber  \\
\left. \times D_{+} \left(\pi\sqrt{\frac{\mbox{Int} (x+1) -x+n}{2K(\gamma_2)M(\gamma_2)}}\right)\right], 
\end{align*}
where $D_{+} (x) = e ^{-x^2} \int _{0} ^{x} e^{t^2} dt$ is the Dawson function. 
The data for the photo-ionization rate $w_{PI}\left(|\vec{E}|\right)$ as a function of the absolute value of the $E$-field and at the fixed value of the electronic
band-gap, $I_p = 9.9$\,eV, are pre-tabulated and used in the propagation code.    

Note that in the current implementation of the model, the Keldysh photo-ionization rate is computed using the instantaneous value of the $E$-field.
This is common in the FDTD Maxwell solver-based treatments of ultrafast laser-matter interactions \cite{popov2011,Bulgakova2013}.
More refined treatment is, in principle, possible through the application of the Yudin-Ivanov formula \cite{Yudin},
but it would require implementing cycle averaging for the $E$-field, which is numerically burdensome.
As we will show below, large variations of the overall pre-factor in the rate formula (\ref{wPI}) do not significantly affect
our results, thus computing the Keldysh rate using the instantaneous $E$-field is justified for our purposes.

\subsection{Avalanche ionization rate}
\label{appendix2}

The velocity of an individual conduction-band electron is a vectorial sum of the time- and position-dependent drift velocity
$\vec{v} _{d}$, which determines the local macroscopic current, and, also vectorial, thermal velocity $\vec{v}_{th}$ of that particular electron. 
Due to the significantly sub-cycle thermalization by the electron-electron collisions, 
the latter has a random direction and the magnitude distributed according to the Maxwell distribution 
with the local temperature $T_e$.
The avalanche ionization rate $w_{AI}$ is computed by averaging the Keldysh impact ionization rate
for a single impactor electron \cite{KeldyshAvalanche} 
over the Maxwell distribution of the thermal velocity component of the conduction-band electrons. 
This formalism has been introduced for the case of negligible drift velocity of the conduction-band electrons 
in \cite{Penano2005} and generalized to the case where the drift velocity is not negligible in \cite{morel2022}.
In the situations we consider, the maximum (over the interaction volume) kinetic energies corresponding to the drift and thermal velocity components are of the same
order ($\sim$10\,eV).

In the spherical coordinate system with the polar axis directed along $\vec{v}_d$ and $\theta$ being the angle between
$\vec{v}_d$ and the variable thermal electron velocity $\vec{v}_{th}$, the explicit expression for the avalanche ionization
rate, averaged over the electron velocity distribution, is as follows:
\begin{eqnarray}
\label{Avalanche} 
w_{AI} = 2 \pi \int_0^\infty v_{th}^2 \cdot \mbox{d} v_{th}
\int_0^\pi \sin \theta \cdot \mbox{d} \theta 
\cdot \Theta \left( W - I_p \right) \nonumber \\ 
\times \alpha_0 \left( \frac{W - I_p}{I_p} \right)^2 
\left( \frac{m_e^*}{3 \pi k_b  T_e} \right)^{3/2}
\exp \left( -\frac{m_e^* v_{th}^2}{3k_b T_e} \right) ,
\end{eqnarray}
where $W \equiv m_e^*v^2/2 = m_e^*{v_d}^2/2 + m_e^*{v_{th}}^2/2 + m_e^*{v_d}{v_{th}}\cos{\theta}$ 
is the electron kinetic energy, composed of the drift and thermal velocity contributions and dependent on the angle
between the drift and thermal velocities, and $\alpha_0 = (e^2/4\pi\epsilon_0)^2m_e^*/n_0^4\hbar^3$.
$\Theta \left( W - I_p \right)$ is the theta function, equal to zero for $W < I_{p}$ and equal to one for $W \geq I_{p}$. 
The integral is solved numerically, and the results for the avalanche rate $w_{AI}(|\vec{v_d}|^2, T_e)$ are pre-tabulated, 
for a fixed electronic bandgap $I_p=9.9$\,eV, and used in the computation of the Maxwell propagator.

The formula (\ref{Avalanche}) reduces to the avalanche ionization rate
derived by Pe\~{n}ano et al. in \cite{Penano2005} for the case of negligible drift velocity 
$\vert \vec{v}_d \vert \ll \sqrt{3 k_B T_e /m_e^*}$. 
The formula (\ref{Avalanche}) is different from the result derived in \cite{morel2022}
for the case when the electron drift velocity is non-negligible, 
but produces the same order-of-magnitude values of the avalanche ionization rate.

\section{Non-paraxial boundary conditions for the field propagator}
\label{boundary}

To set the boundary conditions for the FDTD Maxwell propagator,
we place an auxiliary, linear medium between an ideal (aberration-free), high-NA focusing optic and the sample.
Without loss of generality, the auxiliary medium is assumed to be birefringence-free, with its refractive index $n_0$ matching that of the unperturbed sapphire sample.
The Ignatovsky representation for the real-valued Cartesian components of the electric $\vec{E}(x,y,z)$ and magnetic $\vec{H}(x,y,z)$ 
field vectors of a monochromatic, arbitrarily tightly focused, linearly propagating field everywhere in space
can be expressed as real parts of the following complex quantities
\cite{peatross2017}: 
\begin{eqnarray} 
\lefteqn{}
\label{Fields}
&& \displaystyle{\tilde{E}_x = A\left( I_0+I_2\cos{2\gamma}\right) } , \nonumber \\
&& \displaystyle{\tilde{E}_y = A{I_2}\sin{2\gamma}} , \nonumber \\
&& \displaystyle{\tilde{E}_z = -i A{I_1}\cos{\gamma}} , \nonumber \\
&& \displaystyle{\tilde{H}_x = \sqrt\frac{\epsilon_0}{\mu_0}A{I_2}\sin{2\gamma}} , \nonumber \\
&& \displaystyle{\tilde{H}_y = \sqrt\frac{\epsilon_0}{\mu_0}A\left(I_0-I_2\cos{2\gamma}\right)} , \nonumber \\
&& \displaystyle{\tilde{H}_z = -i \sqrt\frac{\epsilon_0}{\mu_0}A{I_1}\sin{\gamma}} ,
\end{eqnarray}
where $I_0$, $I_1$, and $I_2$ are the integrals detailed below, $A = - ik f \exp{(i k f)} $, 
$f$ is the focal length of the focusing optic in the medium,
$k = 2\pi{n_0}/\lambda _{0}$ is the wave vector in the medium with the index of refraction $n_0$, and
$\lambda _{0}$ is the optical wavelength in vacuum. (See the note below on the treatment of the few-cycle pulses with the optical bandwidth comparable
to their center frequency.) 
The trigonometric functions in (\ref{Fields}) are related to the Cartesian coordinates
$(x,y,z)$ as follows:
$\cos{2\gamma} = (x^2-y^2)/r^2$, $\sin{2\gamma} = 2xy/r^2$, $\cos{\gamma} = {x}/{r}$, and $\sin{\gamma} = {y}/{r}$; $r^2 = x^2 + y^2$.
The origin of the Cartesian coordinate system is set at the geometrical focus of the focusing optic, with the $z$-axis pointing
along the beam axis. The incoming $E$-field is polarized along the $x$-axis.
We assume a perfect refractive-index matching between the focusing optic and the sample, 
thus the Fresnel loss and aberrations upon crossing the entrance boundary of the sample are absent.
This situation represents an experiment with perfect index matching between an aberration-free focusing objective and the sample.

For an axially symmetric, Gaussian input beam, which is the case we consider, $I_{1}$, $I_{2}$, and $I_{3}$
in (\ref{Fields}) are explicitly given by the following integrals over a single variable $\theta$:
\begin{eqnarray} 
\lefteqn{}
\label{Integrals}
&& \displaystyle{I_0 = \int_{0}^{\pi} E_{env} (\rho ') \, \sin{\theta}\, M(\theta) \,
J_0 (k r \sin{\theta}) \, d\theta} , \nonumber \\
&& \displaystyle{I_1 = 2 \int_{0}^{\pi} E_{env} (\rho ') \, \sin{\theta}\, \sqrt{\xi(\theta)} \, M(\theta) \, 
J_1 (k r \sin{\theta}) \, d\theta} , \nonumber \\
&& \displaystyle{I_2 = \int_{0}^{\pi} E_{env} (\rho ') \, \sin{\theta}\, \, \xi(\theta) \, M(\theta) \,
J_2 (k r \sin{\theta} ) \, d\theta} ,
\end{eqnarray}
where 
$E_{env} (\rho ') = E_0\exp{\left[- (\rho ' / {w_0} )^2 \right]}$ is the transverse electric-field distribution of 
the linearly polarized Gaussian beam incident on the focusing optic, $w_{0}$ is its half width at the 
$1/e^2$ intensity level, $\rho ' (\theta) = 2f \sqrt{\xi (\theta)}$, $M(\theta) = \exp{( i k z \cos{\theta} )}$, 
and $\xi(\theta) = (1-\cos\theta)/(1+\cos\theta)$. $E_{0}$ is the normalization factor depending on the input pulse energy and duration. 

The above expressions are strictly valid for a monochromatic field. We assume that the incident laser pulse
has multiple optical cycles, and its spectral width in the frequency domain is much smaller than its center frequency $\omega _{0}$.
Then, the value of the $k$-vector in the above expressions is approximately constant, 
and the temporal dependence of the field factors off from its spatial dependence. 
In time domain, we assume an incident Gaussian pulse with a complex amplitude
$\propto \exp\left[-i\omega _{0}{t}-\left(t/\tau\right)^2\right]$, where
$\tau$ is one half of the pulse duration at the $1/e^{2}$ intensity level. 

For a few-cycle optical pulse with the spectral width comparable to its center frequency, 
the above formulation is valid for the individual temporal Fourier components of the pulse.
In that case, the boundary condition for the nonlinear
Maxwell propagator will have to be set by performing integration of the above expressions (\ref{Fields}) 
over the spectrum of the pulse incident on the focusing optic. 


\begin{figure}[t!]
\includegraphics[width=90mm]{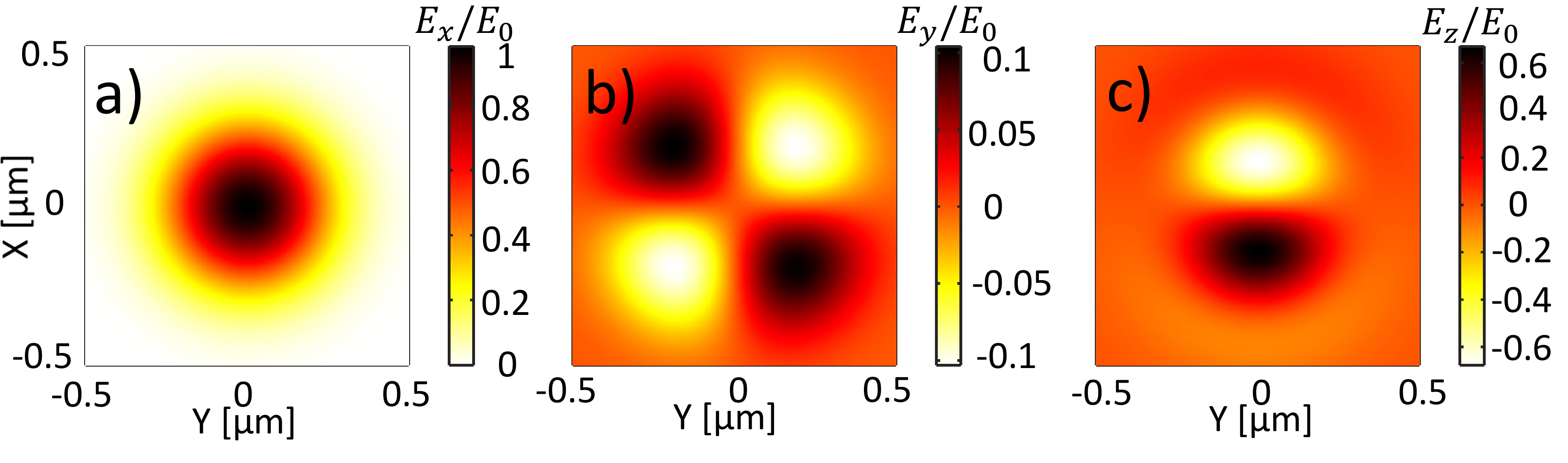}
\caption{Components of the electric field vector
at the focal plane of a parabolic mirror, computed according to (\ref{Fields},\ref{Integrals})
and normalized to the maximum absolute value of the $E$-field at the focus.
The input laser beam at the wavelength of 800\,nm 
is focused into a linear medium with the index of refraction of 1.76. The effective NA of the focusing optic is 1.35.}
\label{fig_1}
\end{figure}

The three real-valued components of the electric field vector in the focal plane of the focusing optic 
are shown in Figure~\ref{fig_1} for the case of linear focusing with NA=1.35.
Note that the longitudinal component $E_z$ of the $E$-field vector near the focus is comparable to $E_x$, its component along the polarization
direction of the incident laser beam. 

To set up the simulation for the nonlinear propagation of the tightly focused laser beam through the interaction volume,
the boundary conditions for the nonlinear Maxwell propagator are set,
using formulas (\ref{Fields},\ref{Integrals}),
some distance before the focal plane, where the nonlinear response of the medium is about to become non-negligible.
If the paraxial boundary conditions were used to initiate the propagator, only the $E_{x}$ and $H_{y}$ components of the 
incident field at the entrance to the computational domain would be specified, leading to erroneous results for the field computed 
by the propagator.

\begin{figure}[b!]
\includegraphics[width=90mm]{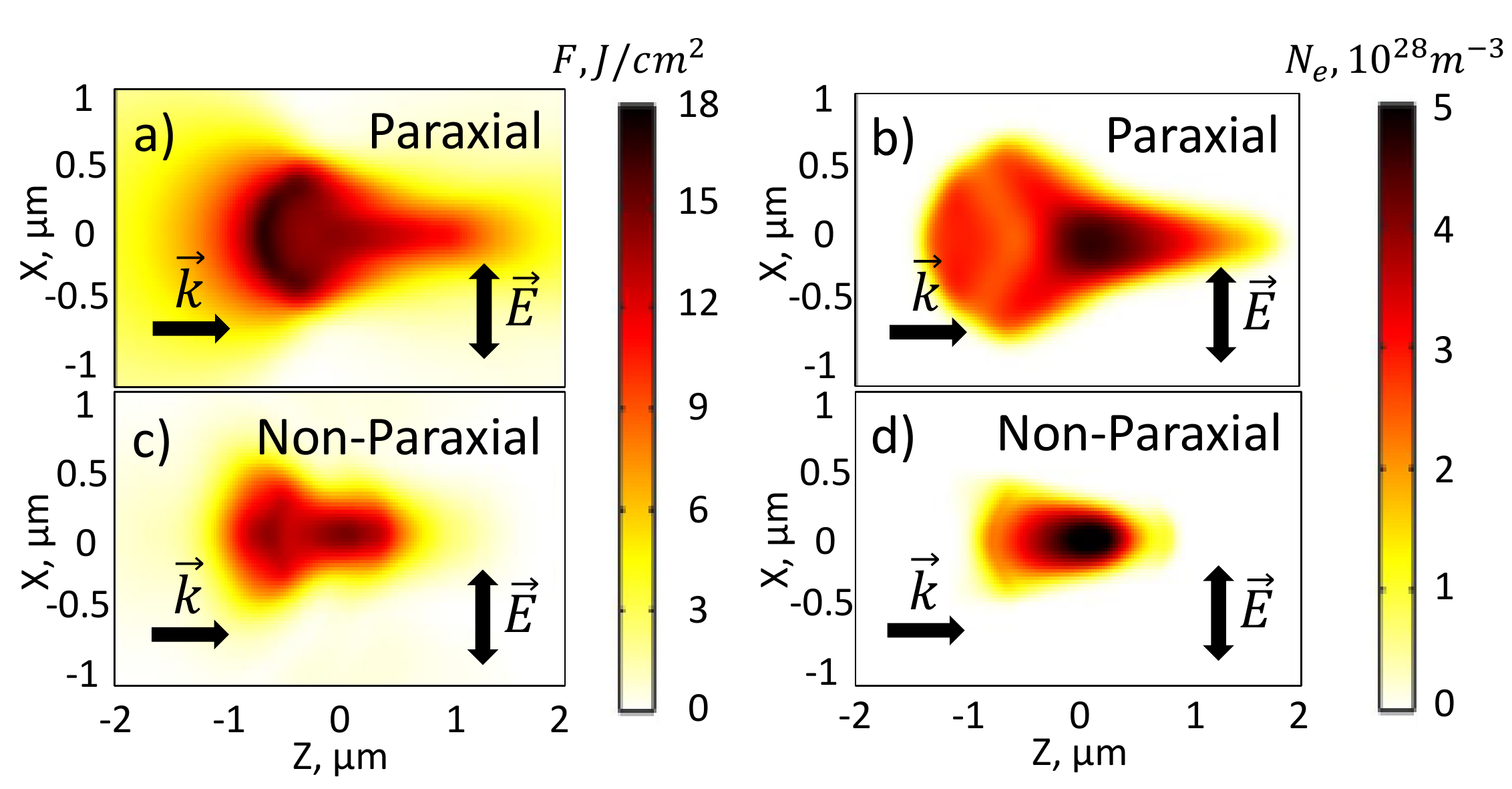}
\caption{\label{fig_2} Spatial distributions of (a,c) the optical fluence and (b,d) the conduction-band electron density
upon the nonlinear pulse propagation in sapphire 
in the cases of (a,b) incorrect, paraxial and (c,d) correct, non-paraxial boundary conditions for the input laser field. 
The energy of the incident laser pulse is 100\,nJ, the FWHM pulse duration is 150\,fs, the effective NA of the focusing optic is 1.35,  
and the material parameters are defined in Section \ref{model} and Appendices.} 
\end{figure}

The importance of using the non-paraxial boundary conditions in the nonlinear simulations with large values of
NA is illustrated by the data shown in Figure~\ref{fig_2}. 
The spatial distributions of the optical fluence and electron density immediately after the passage of the laser pulse through the focal volume are qualitatively different in the data 
obtained using incorrect, paraxial boundary conditions, that specify only the $E_{x}$ and $H_{y}$ components of the source
field at the input boundary of the computational domain, and those obtained using correct, non-paraxial 
boundary conditions detailed above.

Interestingly, the clamped peak values of the fluence and electron density are essentially the same for the simulations
using paraxial and non-paraxial boundary conditions. This is because the origin of the clamping
is in the shielding of the field on its way towards the focus by the generated plasma. The nature of this limiting mechanism
is independent of the specific way the input field is initialized.

\section{Numerical integration scheme and computational aspects}
\label{numerical}

In the FDTD Maxwell propagator, the real-valued electric and magnetic fields are defined on the interleaved square grids shifted with respect to each other by one half
of one sampling interval in all three spatial dimensions. 
At the boundaries of the computational domain, the convolutional perfect matched layers (CPMLs) \cite{Roden2000}, 
with the thickness of 15 computational cells, are set up in order to eliminate the non-physical boundary reflections. 
The auxiliary differential equation (ADE) technique is used to compute the transient optical properties --
the real-valued dielectric permittivity and the polarization, ionization, and Kerr currents
\cite{Greene2006}.

A fixed spatial sampling step $\Delta = 10$ nm is used in all three spatial dimensions $(x,y,z)$, in all simulations with the excitation
at the carrier wavelength of 800\,nm. The chosen resolution is sufficient to properly sample the generated plasma skin layers
and the evanescent fields.
The corresponding temporal step $\delta{t} = \Delta/2c \approx 16.7$ attoseconds 
is chosen to satisfy the Courant-Friedrichs-Lewy 
condition for the stability of the FDTD scheme. 
The optical cycle is well-resolved with about 160 sampling points per cycle.
The computational domain is 6\,$\mu$m-wide in all three spatial dimensions, sampled over 
$600 \times 600 \times 600$ spatial points. The nonlinear propagator runs over a 600\,fs -- long time interval, sampled over
36,000 time steps. The peak of the incident 150\,fs (FWHM) laser pulse is set to reach the boundary 
of the computational domain at 300\,fs. 

In total, eleven equations are solved: Six for the components of the electric and magnetic fields,
three for the components of the average (drift) electron velocity, which defines the electron current density,
and two equations for the electron density and temperature. Additional equations are solved to evaluate the CPMLs.    

For the case of a shorter carrier wavelength (400\,nm), a spatial step $\Delta = 5$\,nm is used, 
while for the excitation at a longer carrier wavelength of 1,500\,nm, we use $\Delta = 20$ nm.
Those two cases are considered in Appendix~\ref{wavelength}.

At each time step, the algorithm updates the components of the magnetic field, 
solves the nonlinear equations for the components of the electric field, then uses the computed fields to update 
the electron current density, number density, and energy density by solving the electron plasma fluid equations complementing Maxwell equations in (\ref{Plasma}). 
The propagator is sourced by adding the incoming non-paraxial input field,
computed according to the procedure detailed in Appendix~\ref{boundary}, to the field computed by the nonlinear propagator at the entrance boundary to the computational domain.

The algorithm is implemented on the graphics processing units (GPUs) card using $10$\,Gb of the GPU memory, 
allowing for the $60 \times$ acceleration relative to the computation with a single processor. 
The completion of a single simulation takes about 
12 hours of computing time.

\section{Impact of the uncertainty of the model parameters on the simulation results}
\label{parameters}

\begin{figure*}[ht!]
\includegraphics[width=150mm]{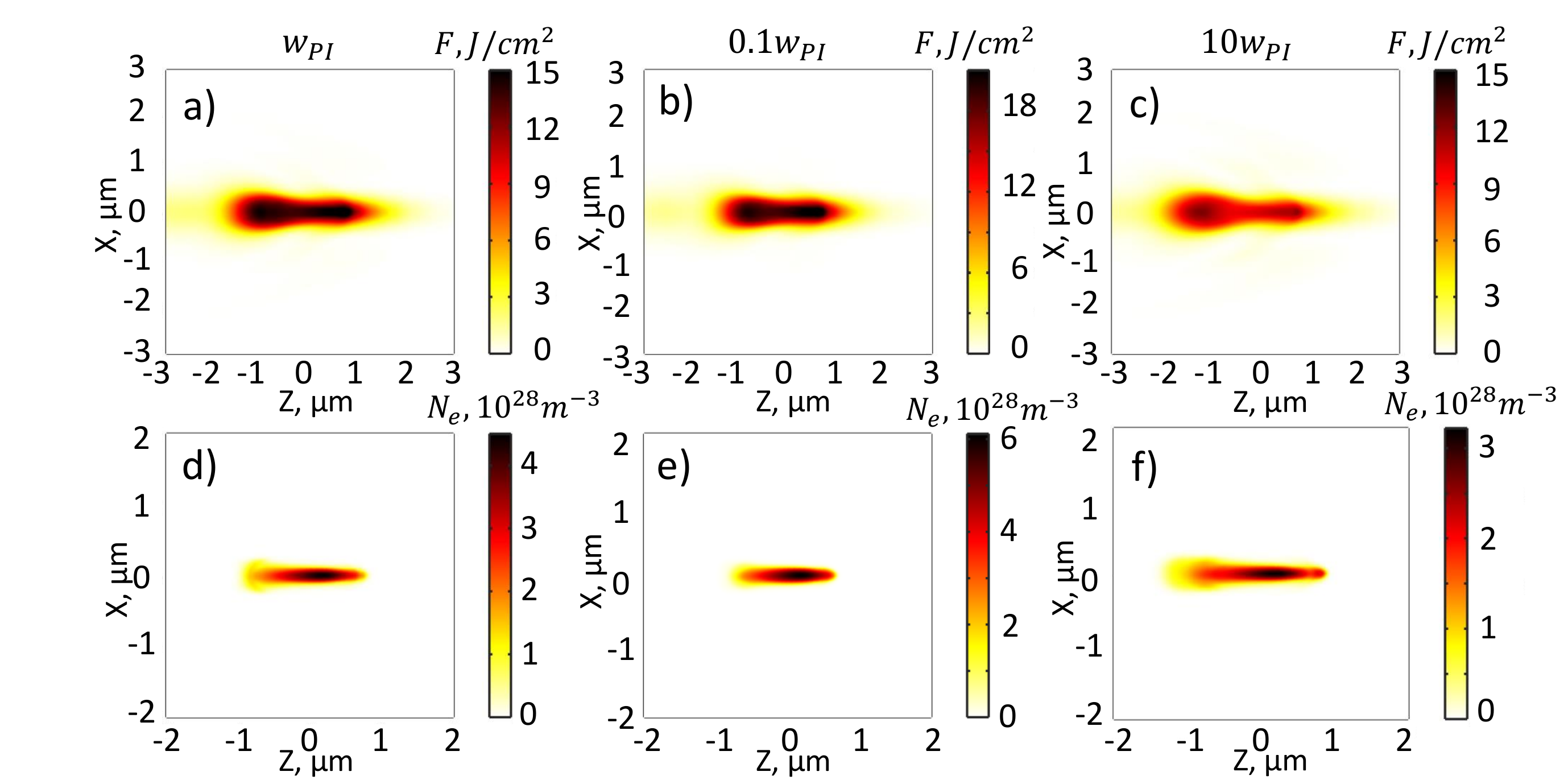}
\caption{\label{fig_3} Spatial distributions of (a-c) the optical fluence and (d-f) the conduction-band electron density for different values of the Keldysh photo-ionization rate: (a,d) $w_{PI}$, the nominal value computed according to (\ref{wPI}), (b,e) $0.1 \times w_{PI}$, and (c,f) $10 \times w_{PI}$. The rest of the material parameters are at their nominal values. The input laser field is the same as that used in Figure \ref{fig_1m}.}
\end{figure*}

As we pointed out earlier, various material parameters that we use in our model
are not precisely known and/or dynamic, i.e., depend on the level of material excitation. Here we show that the main conclusions
based on our simulations, namely the computed values of the peak density of the deposited electromagnetic energy and the peak transient pressure 
in the microexplosion, are not qualitatively dependent on those, potentially uncertain and dynamic, material parameters. 
To that end, we systematically varied those parameters,
one-by-one, by up to one order of magnitude up and down from their nominal values, collected from various literature sources. 
The peak volumetric density of energy deposited by the laser field in the material was found to be limited to $\sim$300\,nJ/$\mu$m$^3$
over the entire range of parameter variations that we considered.
The corresponding maximum transient pressure attainable in the microexplosion is limited to $\sim$250\,GPa, 
which is higher than the value obtained with the nominal parameter values but still
lower than the value reported in \cite{juodkazis2006} ($P_{max} = 13.4$ TPa) by about two orders of magnitude. The peak values of the various quantities computed using our model are summarized in Table~I.

\subsection{Variation of the Keldysh photo-ionization rate}

In our simulations, the instantaneous Keldysh photo-ionization rate $w_{PI}$ is calculated according to Appendix \ref{appendix1}, using fixed values for the electron bandgap of sapphire $I_p = 9.9$\,eV 
and for the effective electron mass $m_e^* = 0.4\,m_e$.  
The Keldysh formulation uses the two parabolic band approximation of the complex electronic structure of the solid, sapphire in this case.
The simulation results obtained using artificially reduced ($0.1 \times w_{PI}$) and increased ($10 \times w_{PI}$) strong-field ionization rates are shown in Figure~\ref{fig_3} 
and summarized in the Table~I.
Such vast variations of the ionization rate result in the modest changes ($\sim 10 \%$) of the peak pressure of sapphire in the WDM state. 

\subsection{Variation of the avalanche ionization rate}

\begin{figure*}[ht!]
\includegraphics[width=150mm]{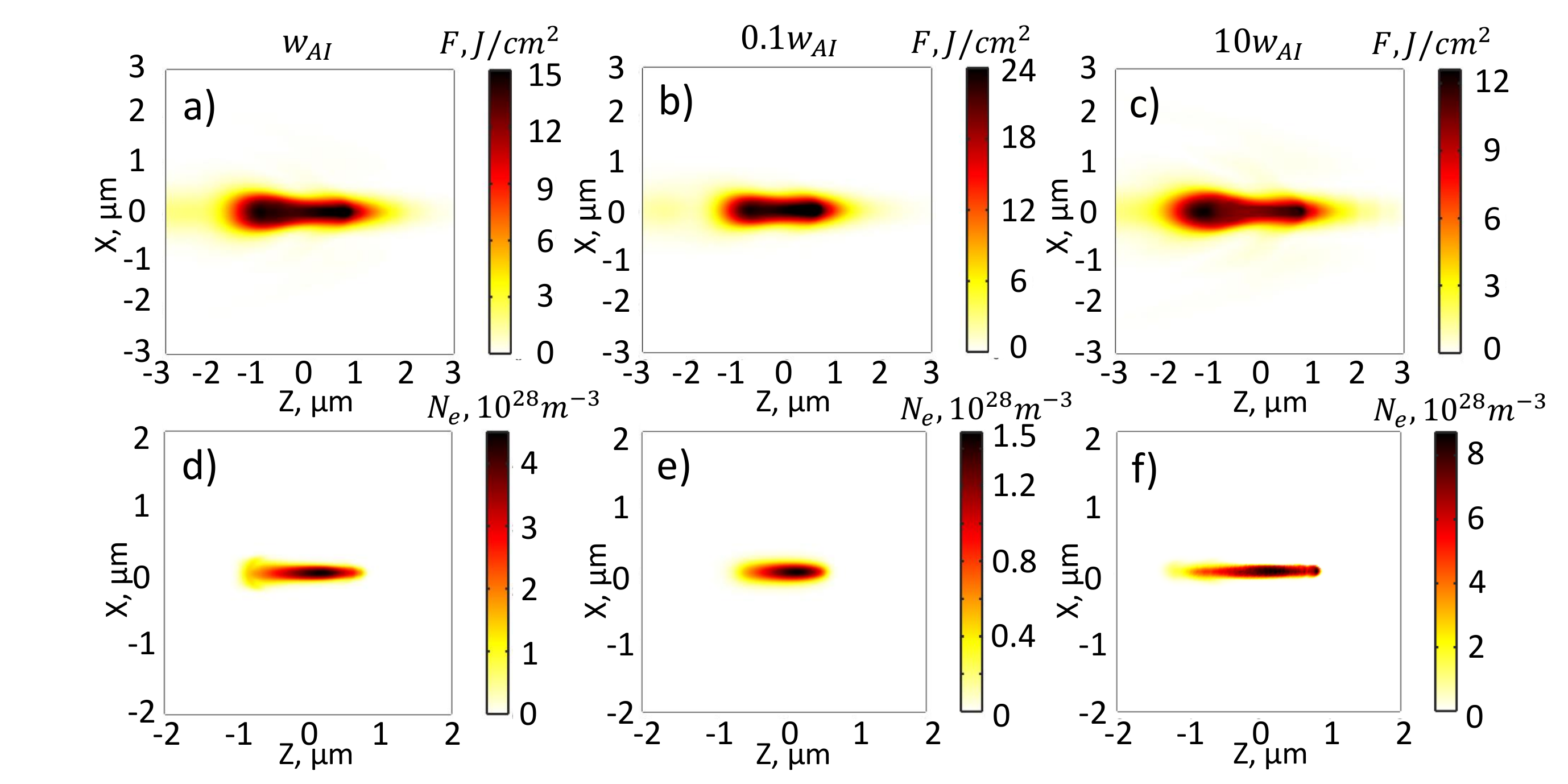}
\caption{\label{fig_4} Spatial distributions of (a-c) the optical fluence and (d-f) the conduction-band electron density for different values of the avalanche ionization rate: (a,d) $w_{AI}$, the nominal value computed according to (\ref{Avalanche}), (b,e) $0.1 \times w_{AI}$, and (c,f) $10 \times w_{AI}$. The rest of the material parameters are at their nominal values. The input laser field is the same as that used in Figure \ref{fig_1m}.}
\end{figure*}

Avalanche ionization multiplies the seed conduction-band electrons produced by Keldysh photo-ionization. In our model, we compute the rate for avalanche ionization 
by averaging the single-electron rate over the shifted Maxwell distribution for the conduction-band electrons, as detailed in Appendix \ref{appendix2}. This formulation may be overly simplified,
as the only material-specific parameters in the expression (\ref{Avalanche}) for the avalanche rate
are the reduced electron mass $m_{e}^*$ and the refractive index $n_0$ of the unperturbed sapphire.
As with the case of the Keldysh photo-ionization rate, we test the dependence of our results on large variations of the pre-factor $\alpha _{0}$ in (\ref{Avalanche}). 
The results of those tests are shown in Figure~\ref{fig_4} and summarized in Table~I. Although the changes of the peak electron density and temperature 
resulting from the variations of
parameter $\alpha _{0}$ are both not insignificant (yet not qualitative), those variations are in opposition to each other, and the resulting change in the peak absorbed energy density is within 30\% at the most, for the case of  the avalanche rate of ten times its nominal value. The corresponding change of the peak transient pressure of sapphire in the WDM state is $\sim 50$\,GPa.

\begin{figure*}[ht!]
\includegraphics[width=150mm]{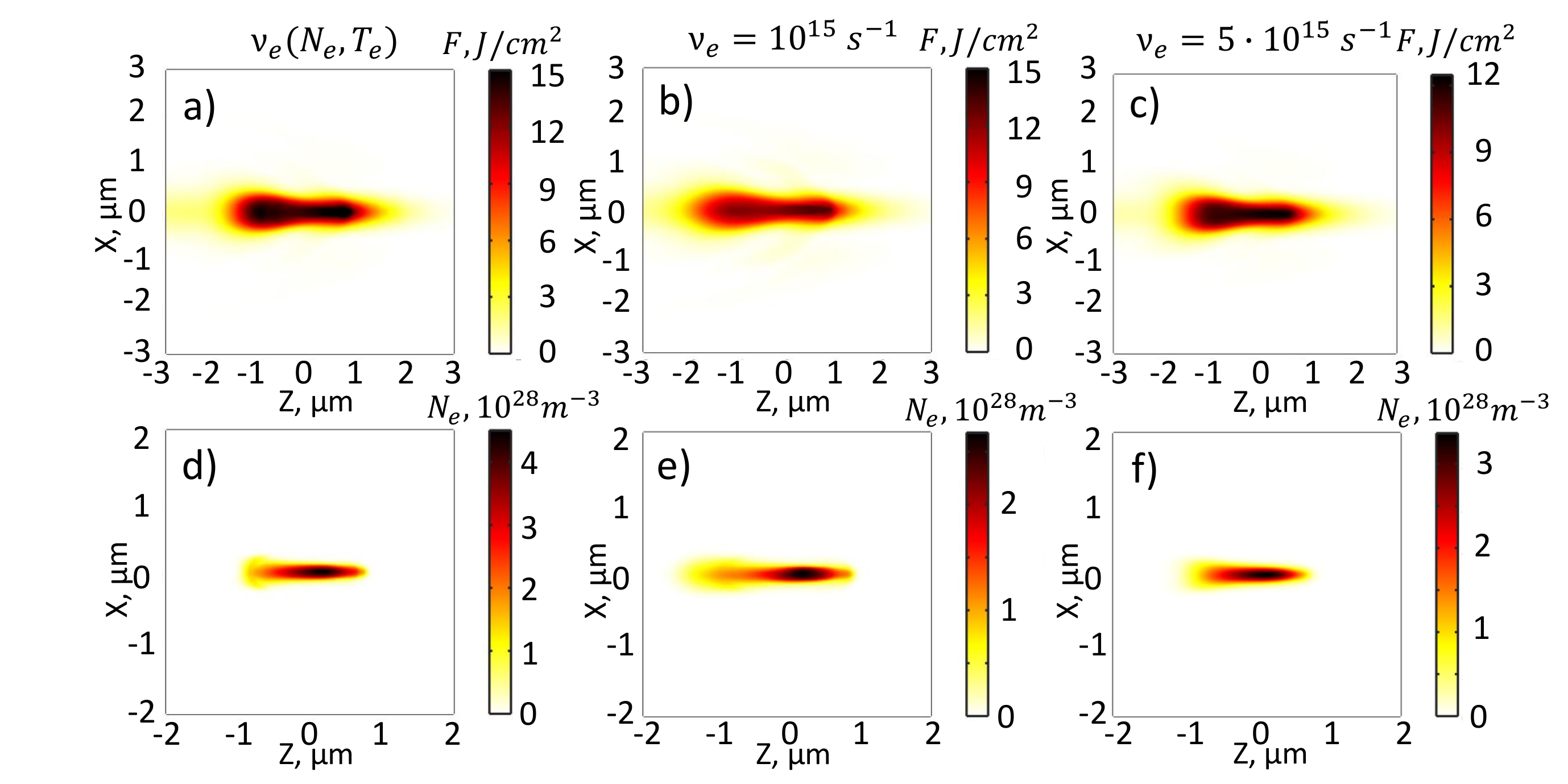}
\caption{\label{fig_5} Spatial distributions of (a-c) the optical fluence and (d-f) the conduction-band electron density 
for different values of the electron collision rate: 
(a,d) the nominal, dynamic $\nu_e(N_e,T_e)$, as detailed in Appendix \ref{appendix3}, (b,e) $\nu_e = 10^{15}$ s$^{-1}$, and (c,f) $\nu_e = 5\cdot{10}^{15}$ s$^{-1}$.
The rest of the material parameters are at their nominal values. The input laser field is the same as that used in Figure \ref{fig_1m}.}
\end{figure*}

\subsection{Variation of the electron collision frequency}

The electron collision frequency $\nu_e$ is an important parameter that determines the transient optical properties of a photo-excited solid. 
As detailed in Appendix \ref{appendix3}, in our model, $\nu_e$ has contributions from the electron-phonon, electron-hole, and electron-ion collisions,
with different contributions dominating over the rest, depending on the level of ionization and plasma heating.
This formulation involves various assumptions that may not be strictly or even qualitatively valid for high values of the conduction-band electron density and temperature.
Here we show that this uncertainty does not qualitatively affect our main conclusions about clamping of the deposited energy density
and the corresponding peak transient pressure in the microexplosion. To validate our conclusions, 
we perform our simulations using two different constant values of $\nu_e$ of $10^{15}$\,s$^{-1}$ and $5 \times 10^{15}$\,s$^{-1}$. 
These values are representative of what have been previously used in literature on the Drude response in photo-excited solids \cite{wu2005,temnov2006,gamaly2014}. 
The results of these tests, summarized in Figure~\ref{fig_5} and in Table~\ref{table1}, show that neither the peak fluence in the interaction zone nor
the peak density of the deposited electromagnetic energy deviate significantly from their baseline values obtained with the comprehensive (dynamic) 
treatment of the electron collisions detailed in Appendix \ref{appendix3}.      

\begin{figure*}[ht!]
\includegraphics[width=150mm]{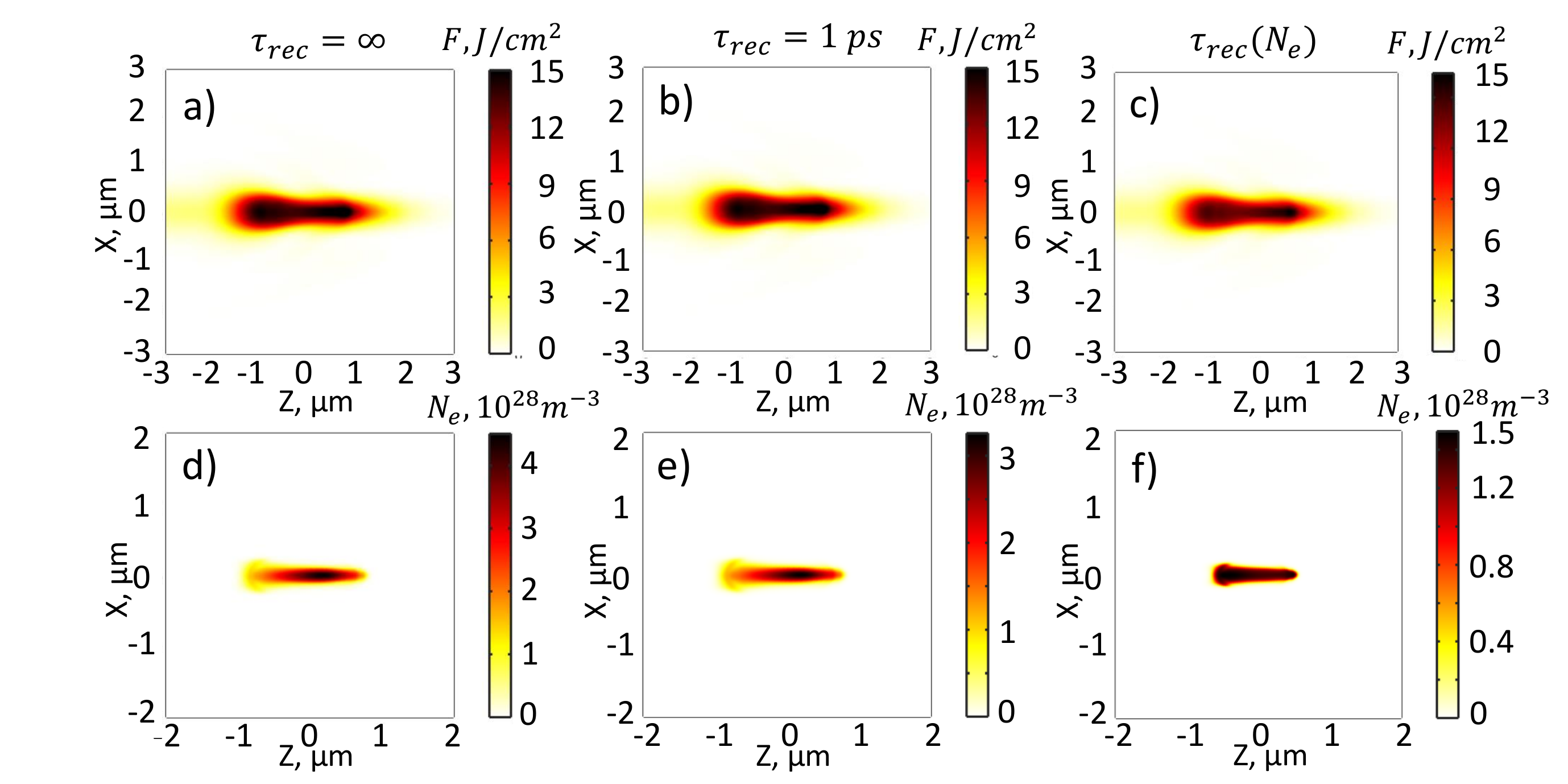}
\caption{\label{fig_6} Spatial distributions of (a-c) the optical fluence and (d-f) the conduction-band electron density: (a,d) neglecting electron recombination, 
(b,e) with linear recombination with the characteristic recombination time $\tau_{rec} = 1$\,ps, 
and (c,f) with a quadratic dependence of the recombination rate on the electron density, such that  $\tau_{rec} = 1$\,fs at the point of complete ionization of the upper valence band of sapphire. 
The rest of the material parameters are at their nominal values. The input laser field is the same as that used in Figure \ref{fig_1m}.}
\end{figure*}

We point out that the computed parameters of the dense electron plasma are consistent with the experimental data on the optical skin depth and reflectivity in the ultrafast surface ablation of sapphire \cite{puerto2010, waedegaard2014, garcia2017}.

\subsection{Accounting for electron recombination}

\begin{figure*}[tb!]
\includegraphics[width=150mm]{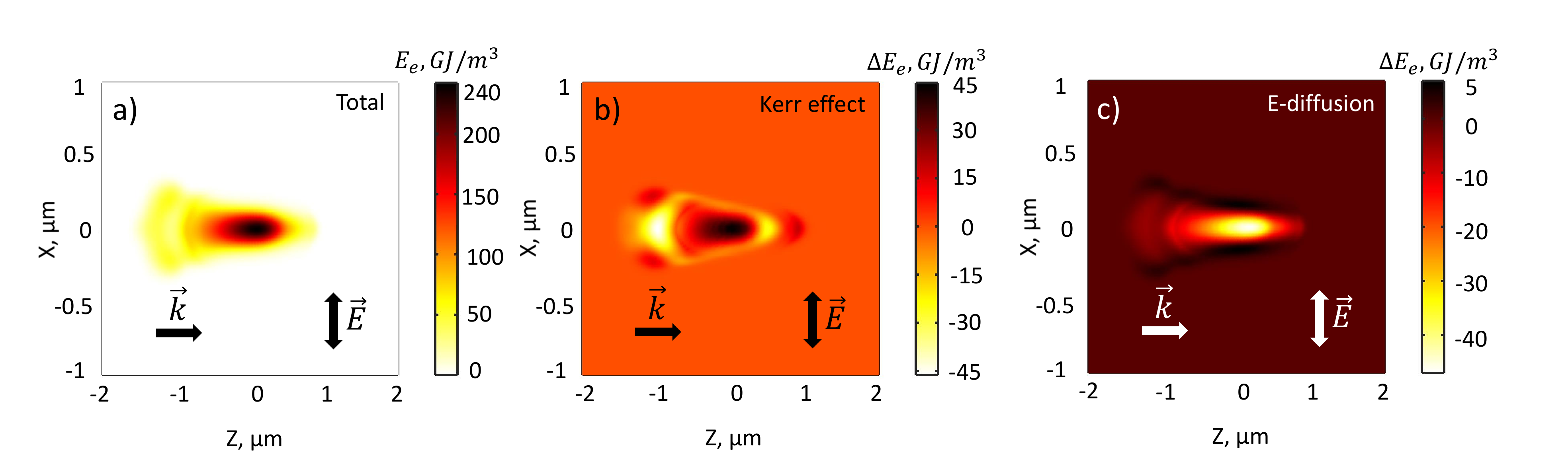}
\caption{Distributions of (a) the total deposited energy density and the contributions to it (b) from the Kerr effect and (c) from the electron diffusion. 
Parameters used in these simulations are the same as those used in Figure~\ref{fig_1m}.}
\label{fig_9}
\end{figure*}  

Electrons excited into the conduction band of a solid can recombine with the accompanying holes. 
Recombination in photo-excited solids is commonly parametrized by the characteristic recombination time $\tau _{rec}$ and quantified, both in
fitting the experimental data and in modeling, by a linear decay term $d N_e / dt \propto - N_e / \tau_{rec}$. The description of recombination, which is a complex, multi-channel process, 
in terms of a single constant parameter $\tau _{rec}$ is certainly an approximation. However, for the purposes of our numerical study, which aims to simulate the photo-excitation phase 
of the laser-material interaction, covering few hundreds of femtoseconds at the longest, the electron-hole recombination can be safely neglected, 
as in sapphire it occurs on a much longer time scale.    
Indeed, the experimental data on the transient surface reflectivity of photo-excited sapphire indicate that the conduction-band electron plasma decays on the time scales
in the range from several to several hundreds of picoseconds \cite{puerto2010, garcia2017}, i.e., one to three orders of magnitude longer than the total simulation time window we consider.
Recombination times on the order of 100\,ps are commonly used in numerical simulations of surface ablation of sapphire \cite{Bulgakova2010}.

Differently from the simple treatment with the linear recombination term mentioned above, 
certain authors treat the electron-hole recombination in photo-excited solids as a three-body process, so that the characteristic recombination time 
has an inverse quadratic dependence on the density of the conduction-band electrons: $\tau_{rec} \propto {N_e}^{-2}$.
This is analogous to the formulations used to describe plasma kinetics in gases and, intuitively, it makes sense, since the presence of both an electron and a hole is needed
for the conduction-band electron to relax into the valence band, and the local densities of electrons and holes are very close.
Based on not entirely transparent arguments, in \cite{gamaly2006}, the characteristic recombination time $\tau _{rec}$ is set to 1\,fs at the point of complete ionization,
i.e., at $N_{e} \approx 4 \times 10^{29}$\, m$^{-3}$. In such a formulation, the inverse quadratic dependence of $\tau _{rec}$ on the electron density dictates the recombination time
to be on the order of 25\,fs for $N_{e} \approx 8 \cdot 10^{28}$\, m$^{-3}$, which is the highest concentration of the conduction-band
electrons that we find over the entire range of the parameter space we consider.

\begin{figure*}[tb!]
\includegraphics[width=150mm]{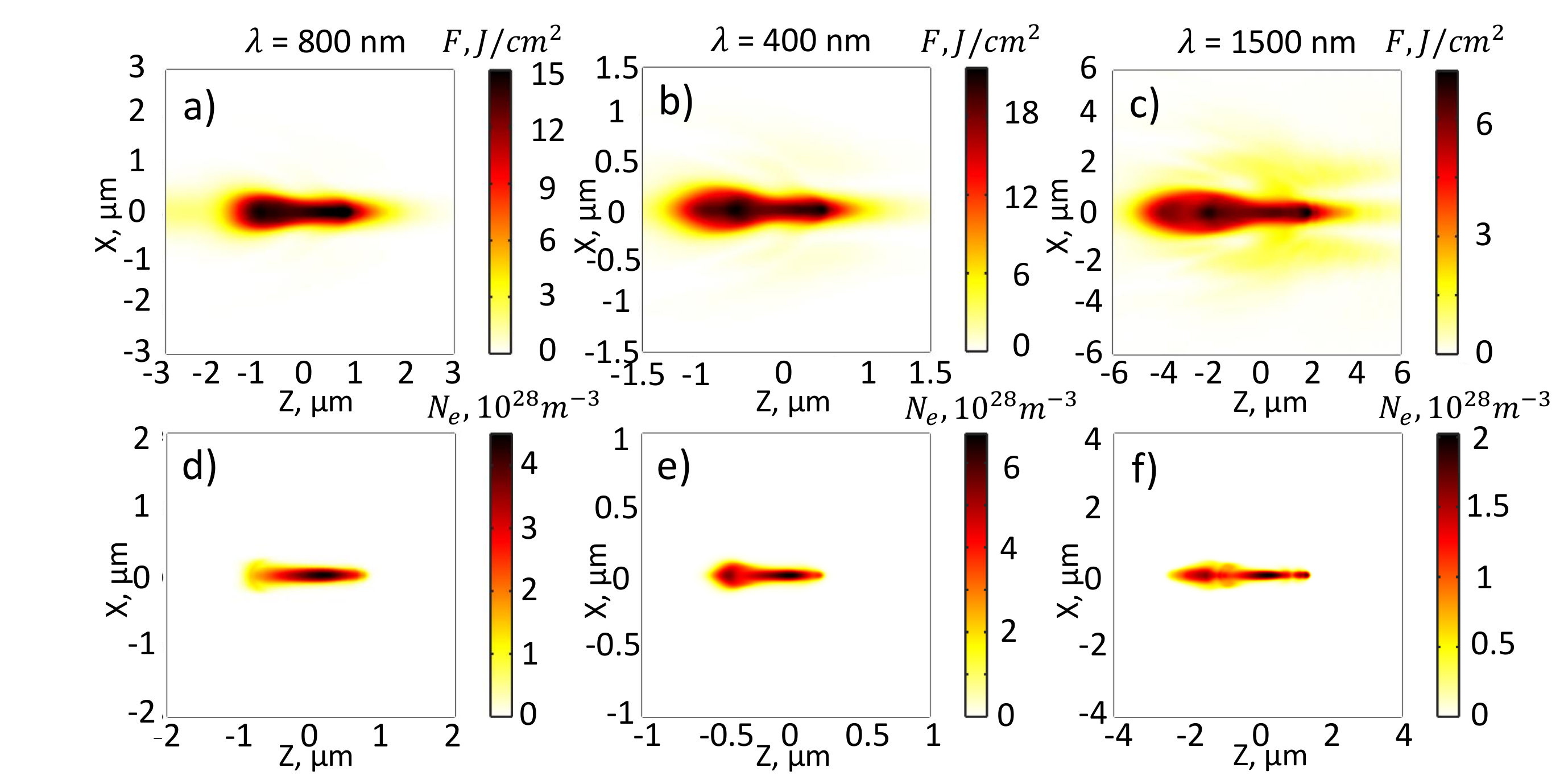}
\caption{Spatial distributions of (a-c) the optical fluence and (d-f) the conduction-band electron density 
for different center wavelengths of the laser pulse:
(a,d) $\lambda _{0} = 800$\,nm, (b,e) $\lambda _{0} = 400$\,nm, and (c,f) $\lambda _{0} = 1,500$\,nm. 
The FWHM pulse duration is 150\,fs in all cases. 
The pulse energy is $100$\,nJ for the cases of $\lambda _{0} = 800$\,nm and $\lambda _{0} = 400$\,nm and $200$\,nJ for $\lambda _{0} = 1,500$\,nm.
The focusing conditions are the same as those used in Figure \ref{fig_1m}.}
\label{fig_10}
\end{figure*}

In this subsection, we show that the recombination of the conduction-band electrons in our simulations can be, indeed, ignored. We test the effect of including two different recombination terms
in the rate equations: a linear term with the constant characteristic decay time $\tau _{rec} = 1$\,ps, which is on the shorter end of the range from the published literature,
and the term, quadratic in the density of the conduction-band electrons, such that the characteristic recombination time is 1\,fs at the point of complete ionization of the valence band.
As evident from the data shown in Figure~\ref{fig_6} and summarized in Table~I, while the maximum fluence over the interaction region is essentially identical in the cases of simulations with and without 
either of the recombination terms, both treatments result in the sizable changes of the peak electron density and temperature.
However, those changes are well within one order of magnitude. As with the other parameter variations, the changes of the peak electron density and temperature are in opposition
to each other, so that the change of the peak deposited energy density, containing contributions from both density and temperature, is even more moderate than
those of the either contribution. For the case of the density-dependent treatment of recombination following \cite{gamaly2006}, ionization stops well before 
the complete depletion of the valence band is reached. Consequently, the rate of recombination never becomes close to the rate of ionization, and the ionization equilibrium scenario,
put forth in \cite{gamaly2006}, is never realized.

\section{Relative roles of different physical phenomena included in the model} 

Ionization and the resulting plasma absorption and refraction are the major physical effects that govern the deposition of the electromagnetic energy in the medium.
In addition to the plasma effects, there are various other linear and nonlinear phenomena that are accounted for in our model. 
It is instructive to examine and quantify the relative importance of those effects.
To that end, we have performed simulations with various effects excluded from the model and compared the results with those obtained using the complete model. 
The results for the two effects contributing in the most significant way are shown in Figure~\ref{fig_9}.
What is shown are the (X,Y) slices of the distributions of the deposited energy, defined as $\mathcal{E}_e = \left[ (3/2){k_B}T_{e} + I_p \right] N_{e} (t=600$\,fs) 
and computed using the complete model,
and of the differences between the complete solution and the solutions obtained with the electron diffusion and the Kerr effects excluded.  
We point out that the values shown in the panels (b) and (c) cannot be considered as the isolated contributions to the deposited energy
by the specific effects in a strict sense, as those contributions are intrinsically coupled. 
However, the data shown in Figure~\ref{fig_9} allow for a semi-quantitative ranking of different phenomena by their importance 
in the overall physical picture of the nonlinear pulse propagation and energy deposition.   

Dropping the Kerr effect from the model results in the reduction of the energy deposition in the center of the focal volume by about 20\%
and in its increase, by about the same amount, in the $\sim$1\,$\mu$m-wide, $\sim$100\,nm-thick shell surrounding the focal volume.
The inclusion of the electron diffusion reduces the deposited energy density in the center of the focal volume by about 20\%, while slightly increasing the energy density
on the periphery of the interaction volume, through the outflow of the hot electrons from the center to the periphery.
The contributions of the convective term and of the magnetic component of the Lorentz force are minor, 
accounting for $\sim$0.15\% and $\sim$0.02\% differences relative to the simulation result using the complete model, respectively.

\section{Simulations with the excitation laser pulses at different carrier wavelengths}
\label{wavelength}

Plasma refraction depends on $\lambda _0$, the center wavelength of the laser pulse, through the $\propto \lambda _{0} ^{-2}$ scaling 
of the critical plasma density.
Therefore, it is expected that using a shorter-wavelength driver could shift the ionization clamping limits, intrinsic to the excitation in the NIR, to higher levels. 
To quantify the dependence of the peak deposited energy and the resulting peak pressure of sapphire in the WDM state on the carrier wavelength of the laser pulse,
we have performed simulations for the laser pulses with spectra centered at 400\,nm and 1,500\,nm wavelengths. 
As for the case of the 800\,nm pulse, the FWHM pulse duration is 150\,fs, and the laser beam is focused into the bulk sapphire 
with the effective numerical aperture NA=1.35.
The dimensions of the simulation volume are scaled with the wavelength of the driver. The spatial and temporal resolutions in the propagator code are adjusted accordingly, 
as specified in Appendix~\ref{numerical}.
The input energy for the case of the 400\,nm pulse is 100\,nJ, which is the same as that for the excitation at 800\,nm, and is increased to 200\,nJ for the 1,500\,nm excitation,
to compensate for the significantly lower multiphoton ionization rate at that wavelength. 

The results of these simulations are shown in Figure~\ref{fig_10}. The peak electron density and temperature, generated by the 400\,nm laser pulse, are $6 \times 10^{28}$\,m$^{-3}$ and $\sim$20\,eV, respectively. As expected, both values are higher than
those obtained with the NIR pulse at the 800\,nm wavelength. However, the peak electron density remains well below the total electron density in the upper valence band
of sapphire. The corresponding peak transient pressure of the WDM sapphire is estimated at 300\,GPa, still significantly short of the multi-TPa levels purported in \cite{juodkazis2006}. 
In principle, the clamping limits can be pushed further up by using a driver pulse with an even shorter center wavelength. For sapphire, the excitation scaling with
wavelength will be limited by the onset of strong linear absorption at around 200\,nm. 

The degree of excitation with the 1,500\,nm driver pulse is predictably lower than what is achieved with the excitation in the NIR and UV: the peak electron density and temperature are $2 \times 10^{28}$\,m$^{-3}$ and $\sim$12\,eV, respectively. The peak transient pressure of the WDM sapphire is $\sim 125$\,GPa.

\bibliography{References}

\begin{flushright}
\end{flushright}

\end{document}